# BENCHMARKING OF NUMERICAL MODELS FOR WAVE OVERTOPPING AT DIKES WITH MILDLY SLOPING SHALLOW FORESHORES: ACCURACY VS. SPEED

1. Christopher H. Lashley [a,*]
2. Barbara Zanuttigh [b]
3. Jeremy D. Bricker [a, c]
4. Jentsje van der Meer [d,e]
5. Corrado Altomare [f, g]
6. Tomohiro Suzuki [a,h]
7. Volker Roeber [i]

a) Dept. of Hydraulic Engineering, Delft University of Technology, Stevinweg 1, 2628 CN Delft, the Netherlands.

b) Dept. Civil, Chemical, Environmental and Materials Engineering, University of Bologna, Via Zamboni, 33, 40126 Bologna BO, Italy.

c) Dept. of Civil and Environmental Engineering, University of Michigan, 2350 Hayward St. Ann Arbor, MI 48109 USA.

d) Water Science and Engineering Department, IHE Delft, Westvest 7, 2611 AX Delft, the Netherlands.

e) Van der Meer Consulting, P.O. Box 11, 8490 AA Akkrum, the Netherlands.

f) Universitat Politecnica de Catalunya – BarcelonaTech, carrer Jordi Girona 1-3, 08034, Barcelona, Spain;

g) Ghent University, Department of Civil Engineering, Technologiepark 60, 9052 Gent, Belgium.

h) Flanders Hydraulics Research, Berchemlei 115, 2140 Antwerp, Belgium.

i) Assistant Professor, E2S chair HPC-Waves Laboratoire SIAME, Pau, France: volker.roeber@univ-pau.fr

*Corresponding author; Telephone: +31684097811; Email: c.h.lashley@tudelft.nl

## Abstract

To accurately predict the consequences of nearshore waves, coastal engineers often employ numerical models. A variety of these models, broadly classified as either phase-resolving or phase-averaged, exist; each with strengths and limitations owing to the physical schematization of processes within them. Models which resolve the vertical flow structure or the full wave spectrum (i.e. sea-swell (SS) and infragravity (IG) waves) are considered more accurate, but also more computationally demanding than those with approximations. Here, we assess the speed-accuracy trade-off of six well-known wave models for overtopping ($\bar{q}$), under shallow foreshore conditions. The results demonstrate that: i) $\bar{q}$ is underestimated by an order of magnitude when IG waves are neglected; ii) using more computationally-demanding models does not guarantee more accurate results; and iii) with empirical corrections to account for IG waves, phase-averaged models like SWAN can perform on par, if not better than, phase-resolving models but with far less computational effort.

## Software availability

- OpenFOAM – developed by OpenCFD Ltd, the software package is freely available from: https://www.openfoam.com/
- SWASH – developed at Delft University of Technology, the model is available freely from: http://swash.sourceforge.net/download/download.htm
- BOSZ – developed at the University of Hawai'i at Manoa, the model is freely available under request from: Volker.roeber@univ-pau.fr
- XBeach Non-hydrostatic and XBeach Surfbeat – developed by IHE Delft, Deltares, Delft University of Technology and the University of Miami, both models are freely available from: https://oss.deltares.nl/web/xbeach/download
- SWAN – developed at Delft University of Technology, the third-generation wave model is freely available from: http://swanmodel.sourceforge.net/download/download.htm

# 1 Introduction

## 1.1 Background

Coastal engineers often employ numerical modelling in the design, assessment and rehabilitation of coastal structures to accurately forecast nearshore waves and currents, sometimes including the consequences (Akbar and Aliabadi, 2013, Sierra, et al., 2010, Smith, et al., 2012, Suzuki, et al., 2017). Of particular interest is the extent to which waves reach and pass over the crest of a structure, referred to as wave overtopping. Extreme overtopping events are characterized by considerable flow velocities which impose serious hazards to both people and infrastructure; with flooding or coastal inundation as the most critical consequence. The integration of numerical modelling in estimating wave overtopping and the design of coastal structures is becoming increasingly more attractive given the progress in available computing power and the limitations of traditional empirical approaches which are typically limited to the number of simplified structure configurations and the range of environmental conditions applied in their derivation. Furthermore, as many of the empirical models (e.g. EurOtop, (2018)) require the incident significant wave height ($H_{m0,T,toe,in}$) and spectral wave period ($T_{m-1,0,toe,in}$) at the toe of the structure as input, numerical models are often needed to accurately capture the nonlinear effects associated with the shoaling and breaking of high-frequency sea-swell (SS) waves in shallow water (Altomare, et al., 2016, Mase, et al., 2013). Such effects include a rise in mean water level—known as wave-induced setup—and the growth of low-frequency infragravity (IG) waves (Figure 1) which not only contribute to $H_{m0,T,toe,in}$ but also result in higher values of $T_{m-1,0,toe,in}$ (Hofland, et al., 2017).

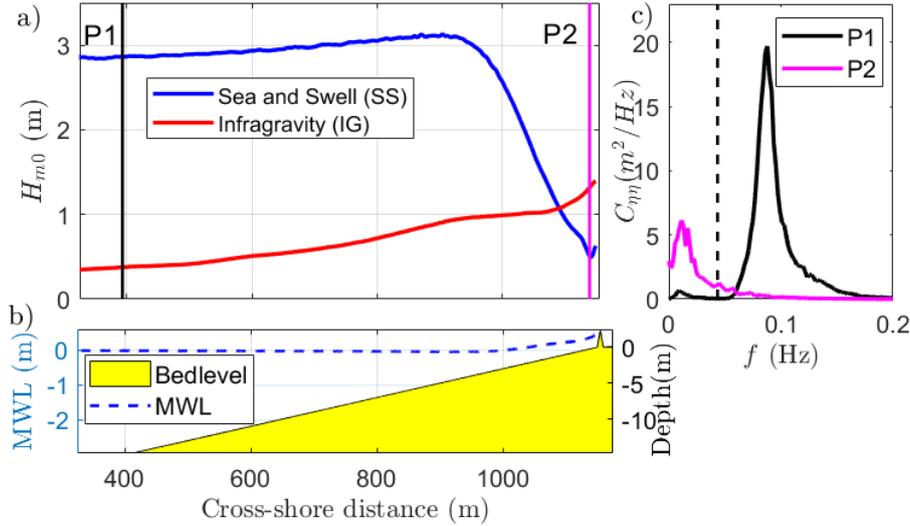

**Figure 1** Schematic representation of wave transformation over a shallow foreshore (from an XBeach model simulation), showing a) the growth of IG waves; b) the increase MWL at the dike toe; and c) the shift in the peak in energy density to lower frequencies from offshore (P1) to the dike toe (P2). Vertical line in panel 'c' indicates the separation between SS and IG frequencies.

A variety of numerical models, which may be broadly classified as phase-resolving or phase-averaged, have been developed for such applications; each with strengths and limitations owing to the physical parameterization of processes and the numerical schemes incorporated within them (Cavaleri, et al., 2007, Vyzikas and Greaves, 2018). Models which attempt to resolve the vertical flow structure and those that consider the full frequency range of nearshore waves (i.e. both SS and IG waves) are considered not only more accurate, but also more computationally demanding than those which make use of approximations.

Within the phase-resolving class of wave models, those that resolve the vertical flow structure and solve the fully nonlinear, time-averaged Navier-Stokes (NS) equations—often referred to as Computational Fluid Dynamics (CFD) or depth-resolving models—have the least theoretical limitations and are generally considered the most accurate. CFD models, such as the mesh-based Eulerian approach OpenFOAM (Jasak, et al., 2007) or mesh-less Lagrangian approach DualSPHysics (Crespo, et al., 2015), are able to simulate complex wave problems, such as: nonbreaking and breaking waves, wave-current interaction and wave-structure interaction from deep to shallow water conditions, including the overturning (Lowe, et al., 2019) and roller formation of breaking waves (Higuera, et al., 2013). However, these models require a significant amount of computational effort (unless a coupling method is applied (Altomare, et al., 2015, Altomare, et al., 2018, Verbrugghe, et al., 2018)); thus, limiting their application so far to very local phenomena—for example, wave overtopping.

As depth-resolved (fully 3D or 2DV) models are generally considered too computationally expensive for operational use, the problem may be further simplified by depth-averaging. These models, in which the vertical structure is not directly resolved but only modelled parametrically, are referred to as two-dimensional in the horizontal (2DH), or 1DH where only a cross-shore transect is simulated (Brocchini and Dodd, 2008). As a result of depth-averaging, processes such as wave overturning, air-entrainment and wave generated turbulence are not directly solved. Those that simulate the amplitude and phase variation of SS waves are often referred to as phase-resolving. Within this type of model, there are generally two main sets of governing equations: i) the Non-linear shallow water (NLSW) equations; and ii) the Boussinesq type.

While the Boussinesq-type models (e.g. FUNWAVE (Kirby, et al., 1998), MIKE21-BOUSS (Warren and Bach, 1992) and BOSZ (Roeber and Cheung, 2012)) directly account for the dispersive properties of waves in deeper water, the NLSW models assume that waves are non-dispersive and are therefore limited to shallow-water applications (Brocchini and Dodd, 2008, Zijlema and Stelling, 2008). This limitation can be removed by taking a SS-wave averaged approach; however, at the cost of decreased accuracy (due to exclusion of SS-wave motions). The high-frequency waves are averaged, resulting in only motions at the scale of the wave group; thus, reducing the computational demand (e.g. XBeach Surfbeat (XB-SB) model (Roelvink and Costas, 2019, Roelvink, et al., 2009)).

In order to use the NSLW equations for phase-resolving simulation of SS-wave motions, Stelling and Zijlema (2003) proposed another method to account for dispersion (a result of non-hydrostatic pressure) whereby the pressure is decomposed into non-hydrostatic and hydrostatic pressure components (e.g. SWASH (Zijlema, et al., 2011), NHWAVE (Ma, et al., 2012) and XBeach Non-hydrostatic (XB-NH) (Smit, et al., 2010) numerical models). This approach improves the dispersive properties without neglecting the higher-frequency motions; however, at the expense of more computational demand. The accuracy and range of applicability of the non-hydrostatic models may be further enhanced by coarsely dividing the model domain into a fixed number of vertical layers ($K \leq 3$); thereby, improving the frequency dispersion (e.g. SWASH, NHWAVE or XB-NH in multi-layered mode (De Ridder, 2018)). By further increasing the number of vertical layers ($K \geq 10$), models like SWASH may be extended to the depth-resolving class. This approach increases the computational demand but allows processes, such as undertow and the shoreward flow near the surface, to be resolved.

Given that phase-resolving models require a grid resolution high enough to resolve the individual SS-wave components, they are generally computationally feasible only for areas of limited size. For large-scale modelling of wave motion, a phase-averaged approach is most commonly used. This type of model is constructed on the assumption that a random sea-state is composed of a superposition of linear waves whose height is a function of their frequency and direction of propagation. For an individual wave train the rate of change of wave energy (or action) flux is balanced by the wave energy transfer among different wave components in different directions and different frequencies, as well as energy input and dissipation. With the phase information filtered out, these models can use much courser computational grids and therefore be applied to large areas. However, as individual waves are not resolved, these models must be combined with empirical formulae to estimate wave run-up and overtopping (Oosterlo, et al., 2018, Sierra, et al., 2010). Commonly used spectral models in nearshore applications include SWAN (Booij, et al., 1999) and STWAVE (Smith, et al., 2001). These models are generally able to accurately reproduce higher harmonics (SS waves); however, they do not account for the interactions that force IG-wave motions (Cavaleri, et al., 2007), which tend to dominate in shallow water.

With respect to previous model comparisons in shallow coastal environments, Buckley, et al. (2014) assessed the performance of SWASH, SWAN and XB-SB in predicting SS wave heights ($H_{m0,SS}$), IG wave heights ($H_{m0,IG}$) and setup ($\bar{\eta}$) across a steep laboratory fringing reef profile (varying from 1:5 to 1:18.8). Results showed that each model was capable of accurately predicting $H_{m0,SS}$; however, SWAN failed to simulate the transformation of energy to lower frequencies and thus, failed to predict $H_{m0,IG}$. Likewise, SWAN showed considerably more error in its prediction of $\bar{\eta}$ compared to SWASH and XB-SB. On the other hand, XB-SB performed comparably well to its phase-resolving counterparts in the prediction of nearshore wave heights; and surprisingly the extent of wave run-up, particularly when IG-waves dominated at the shoreline (Lashley, et al., 2018). From these previous studies, the points of discussion that naturally arise are:

i) Can phase-averaged models like SWAN be accurately applied under very shallow conditions, where IG waves dominate and $\bar{\eta}$ is significant?
ii) Given that IG waves dominate, are models of increasing complexity needed or is a short-wave averaged but IG-wave resolving approach all that is required? and
iii) While attempts at model comparisons for wave overtopping have been made (St-Germain, et al., 2014, Vanneste, et al., 2014), no study to date has the full range of model complexity (from depth-resolving to phase-averaged) or successfully quantified the accuracy versus speed of these models under irregular wave forcing .

## 1.2 Objective

In the present study, it is our primary aim to quantify the accuracy versus speed of computation of six commonly-used nearshore wave models (Table 1) in their prediction of irregular wave overtopping of a dike with very shallow foreshore conditions—where IG waves and setup contribute significantly.

## 1.3 Outline

This report is organized as follows: Section 2 provides descriptions of the physical and numerical models applied, followed by descriptions of key parameters and empirical formulae used in the analysis. It ends with a description of the metrics used to quantify model accuracy. In Section 3, the results of the model-data comparisons and the overall influence of IG waves on overtopping are presented and discussed. Section 4 concludes the report by summarising the findings, acknowledging the present study's limitations and identifying areas for future work.

Table 1 Overview of the numerical models considered for comparative analysis.

| Model | Model Type | | Wave Propagation | | Overtopping |
|---|---|---|---|---|---|
| | | | SS Waves | IG Waves | |
| OpenFOAM | Phase-resolving | Depth-resolving | Directly | | Directly |
| SWASH[a] | | | | | |
| BOSZ | | Depth-averaged | | | |
| XB-NH | | | | | |
| XB-SB | IG-wave resolving; SS-wave averaged | | Action-balance | Directly | Directly for IG waves[b] |
| SWAN | Phase-averaged | | Action-balance | Excluded | Empirically |

[a]Does not resolve wave overturning or wave roller formation.

[b]Does not include SS-wave overtopping.

## 2 Methods

This section begins with a description of the physical model tests under consideration. After which it describes the five numerical models under evaluation, including their governing equations and setup details. A description on the parameters and metrics used to assess model accuracy and computation speed is then provided. Finally, the additional numerical simulations for comparative analysis are described.

### 2.1 Description of the Physical Models

In the present study, we consider two specific test cases that were both performed at Flanders Hydraulics Research in a smooth, 1-m wide section of their 70-m long and 1.45-m deep wave flume (Altomare, et al., 2016) with different deep water wave heights ($H_{m0,T,deep}$), peak periods ($T_p$), foreshore slopes ($m$), initial water depths at the toe ($h_{toe}$), dike slopes ($\alpha$) and dike freeboards ($R_c$) (Table 2). These cases were selected to cover a wide range of deep-water wave steepness ($s_0$), from very mild ($s_0 = 0.007$, typical of swell conditions) to very steep ($s_0 = 0.047$, typical of wind-sea conditions). With relative water depths ($h_{toe}/H_{m0,T,deep}$) < 1, these conditions are considered very shallow (Hofland, et al., 2017). Both experiments simulated irregular spilling waves (with breaker parameter based on $m$, $\xi_{0,fore} < 0.5$) with a duration approximately equal to 500 waves to obtain accurate and comparable estimates of the mean overtopping discharge ($\bar{q}$) (Romano, et al., 2015).

Table 2 Summary of test conditions for both the mild- and steep-wave cases.

| Case | $H_{m0,T,deep}$ (m) | $T_p$ (s) | $\cot m$ | $s_0$ | $kh$ | $h_{toe}$ (m) | $\dfrac{h_{toe}}{H_{m0,deep}}$ | $\cot \alpha$ | $R_c$ (m) |
|---|---|---|---|---|---|---|---|---|---|
| Mild swell | 0.06 | 2.29 | 50 | 0.007 | 0.98 | 0.032 | 0.53 | 2 | 0.06 |
| Steep wind-wave | 0.21 | 1.70 | 35 | 0.047 | 1.45 | 0.025 | 0.12 | 3 | 0.08 |

For the mild swell-wave case, the variations of water-surface elevations were measured using 10 resistance-type gauges, all synchronously sampling at 50 Hz (Figure 2a); while 6 gauges with a sample frequency of 20 Hz were used in the steep-wave case (Figure 2b). In the analysis to follow, the term "offshore" is used to refer to gauges 1 to 7 and 1 to 3 of the mild swell and steep-wind wave cases, respectively; and the term "nearshore" to refer to gauges 8 to 10 and 4 to 6, respectively. In either case, the term "toe" refers to the last wave gauge (gauge 10 and gauge 6 of the mild swell and steep wind-wave cases, respectively).

In both cases, the instantaneous overtopping was measured using two Balluff "Micropulse" water sensors situated inside the overtopping box; and $\bar{q}$ was then obtained by dividing the total volume of water collected at the end during the test by the total test duration.

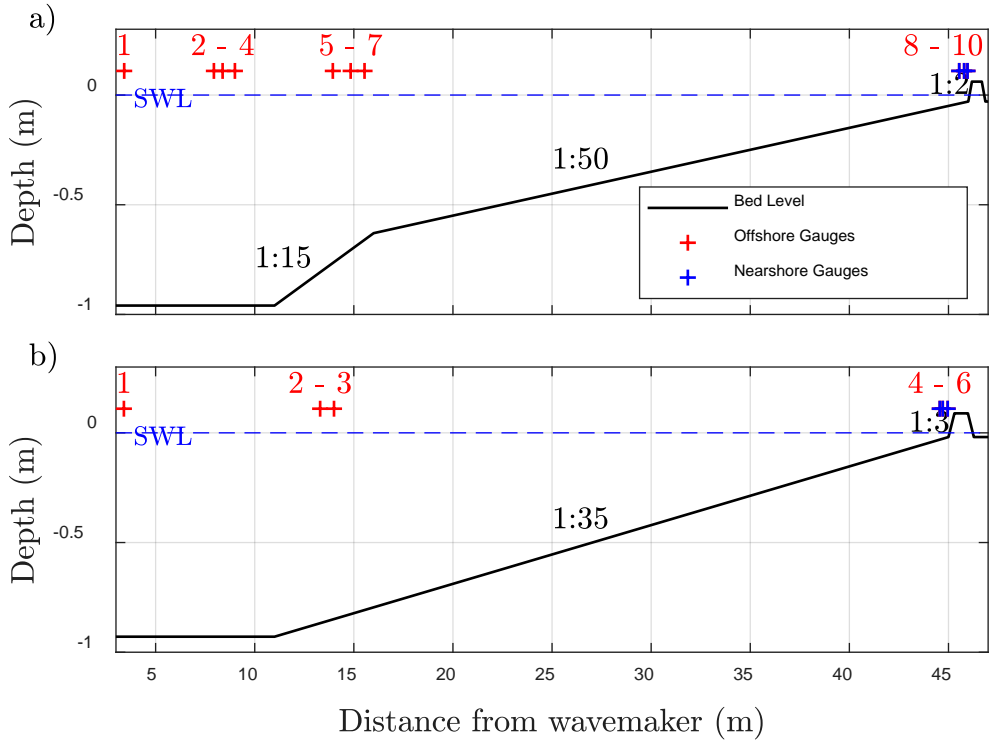

Figure 2 Physical model setups showing gauge locations for both the: a) mild swell; and b) steep wind-wave cases.

## 2.2 Description of Numerical Models

In this study, six widely-used open-source numerical wave models are considered for comparative analysis. Each model is forced at its boundary with still water levels and parametric spectra (JONSWAP) to match those observed at the most offshore wave gauge during each physical experiment. Likewise, the smooth flume bottom was represented as either a Manning coefficient ($n$) of 0.01 s/m$^{1/3}$ or a Nikuradse geometrical roughness ($k_s$) of 0.3 x 10$^{-3}$ m (in the case of SWAN). A general description of each model is provided in the sections that follow. As we investigate two extremes: very mild swell and very steep wind waves, it is reasonable that some calibration was required for the depth-averaged models (BOSZ, XB-NH and XB-SB). Therefore, a description of the main calibration parameters, their optimum values and impact on model results is also provided. In general, calibration was aimed at reducing the error in $\bar{\eta}$ and $H_{m0,T}$.

### 2.2.1 OpenFOAM

The software OpenFOAM is an Open Source object-oriented library, composed by solvers and utilities (Jasak, et al., 2007). The formers are designed to numerically solve continuum mechanics problems, while the latter perform tasks involving data manipulation.

For the present study, the library *waves2Foam*, a toolbox capable of generating and absorbing free surface water waves, has been adopted. Currently, the method applies the relaxation zone technique (active sponge layers) and supports a large range of wave theories (Jacobsen et al., 2012). The governing equations for the combined flow of air and water are given by the Reynolds Averaged Navier–Stokes equations (Equations 1 and 2):

$$\frac{\partial \rho u}{\partial t} + \nabla \cdot [\rho u u^T] = -\nabla p^* - g \cdot x \nabla \rho + \nabla \cdot [\mu \nabla u + \rho \tau] + \sigma_T \kappa_\gamma \nabla \gamma, \tag{1}$$

coupled with the continuity equation (2) for incompressible flow:

$$\nabla \cdot u = 0, \tag{2}$$

where $u$ is the velocity field, $p^*$ is the dynamic pressure component, $\rho$ is the density, $g$ is the acceleration due to gravity and $\mu$ is the dynamic molecular viscosity. The Reynolds stress tensor $\tau$ is defined as:

$$\tau = \frac{2}{\rho} \mu_t S - \frac{2}{3} kI, \tag{3}$$

where $\mu_t$ is the dynamic eddy viscosity, $S$ is the strain rate tensor, $k$ is the turbulent kinetic energy per unit mass and $I$ is the identity matrix. The last term in Equation 1 is the effect of surface tension, where $\sigma_T$ is the surface tension coefficient and $\kappa_\gamma$ is the surface curvature (Jacobsen, et al., 2012). The track of the free surface is performed by using the VOF method (Hirt and Nichols, 1981).

For the mild and the steep cases, two regular slightly different meshes have been generated, to account for the differences between the two wave conditions. The numerical domains of the mild and steep cases are respectively composed by 49021 and by 70316 cells, with a graded mesh both in the *x* (0.3-0.005 m for the mild, 0.1-0.01 m for the steep) and in the *y* (0.05-0.005 m for the mild, 0.1-0.01 m for the steep) directions. In both cases, the selected regular and constant mesh allowed for a fair compromise between the computational effort and the accuracy of the results.

### 2.2.2 SWASH

SWASH is a time domain model for simulating non-hydrostatic, free-surface and rotational flow. It solves the NLSW equations with an added non-hydrostatic pressure correction term (Smit, et al., 2013):

$$\frac{\partial \eta}{\partial t} + \frac{\partial uh}{\partial x} = 0, \tag{4}$$

$$\frac{\partial u}{\partial t} + \frac{\partial uu}{\partial x} + \frac{\partial wu}{\partial z} = -\frac{1}{\rho}\frac{\partial (p_h + p_{nh})}{\partial x} + \frac{\partial \tau_{xx}}{\partial x} + \frac{\partial \tau_{xz}}{\partial z}, \tag{5}$$

$$\frac{\partial w}{\partial t} + \frac{\partial uw}{\partial x} + \frac{\partial ww}{\partial z} = -\frac{1}{\rho}\frac{\partial p_{nh}}{\partial z} + \frac{\partial \tau_{zz}}{\partial z} + \frac{\partial \tau_{zx}}{\partial x}, \tag{6}$$

$$\frac{\partial u}{\partial x} + \frac{\partial w}{\partial z} = 0, \tag{7}$$

where $\eta$ is the free surface elevation; $u(x,z,t)$ and $w(x,z,t)$ are the horizontal and vertical velocities, respectively; $h$ is the water depth; $\rho$ is the density of water; $p_h$ and $p_{nh}$ are the hydrostatic and non-hydrostatic pressures, respectively; and $\tau_{xx}$, $\tau_{xz}$, $\tau_{zz}$ and $\tau_{zx}$ are the turbulent stresses.

The model exhibits good linear dispersion up to $kh \approx 8$ and $kh \approx 16$ with two and three equidistant (sigma) vertical layers ($K$), respectively; its frequency dispersion is further improved by increasing $K$.

Here, the model was applied with $K = 20$, which is sufficient for the phase velocity at the breaking wave front to be computed accurately. As such, no additional control is required to initiate or terminate wave breaking. The vertical pressure gradient was discretized by the standard central differencing scheme with the ILU pre-conditioner. The standard k-ε turbulence model is applied to take into account vertical mixing.

A cross-shore grid spacing ($\Delta x$) of 0.04 m was specified for both the mild- and steep-wave cases. This resulted in approximately 200 and 110 grid cells per deep-water wavelength ($L_0/\Delta x$) for the mild- and steep-wave cases, respectively. For phase-resolving models, $L_0/\Delta x$ is typically kept between 50 and 100 (by rule of thumb) to ensure that the wave components are accurately resolved; however, as waves propagate in very shallow water, the local wavelength becomes much shorter than $L_0$. Thus, in order to maintain a reasonable number of grid cells per local wave length, these higher-than-typical grid resolutions ($L_0/\Delta x = 200$ and 110) were specified.

Rijnsdorp, et al. (2017) proposed a sub-grid approach to improve model efficiency, where vertical accelerations and non-hydrostatic pressures are resolved on a relative course grid while the horizontal velocities and turbulent stresses are resolved on a much finer sub-grid. This approach was attempted here, however, the simulations failed due to instabilities.

### 2.2.3 BOSZ

The BOSZ wave model—which is freely-available upon request from the developers—computes hazardous free surface flow problems ranging from near-field tsunamis to extreme swell ranges generated by hurricanes. It solves the following re-formulated, depth-integrated Boussinesq equations of Nwogu (1993), in vector notation:

$$\eta_t + \nabla[(h+\eta)U] + \nabla \cdot \left\{ \left( \frac{\bar{z}_\alpha^2}{2} - \frac{h^2}{6} \right) h\nabla(\nabla \cdot U) + \left( \bar{z}_\alpha + \frac{h}{2} \right) h\nabla[\nabla \cdot (hU)] \right\} = 0 \quad (8)$$

$$U_t + U(\nabla \cdot U) + g\nabla\eta + \left\{ \frac{\bar{z}_\alpha^2}{2} \nabla(\nabla \cdot U) + \bar{z}_\alpha \nabla[\nabla \cdot (hU)] \right\}_t = 0, \quad (9)$$

where U is the horizontal flow velocity defined at a reference depth $\bar{z}_\alpha = -0.55502h$ (Simarro, et al., 2013).

The governing equations exhibit good dispersion accuracy up to $kh \approx \pi$. Given the difficulty of Boussinesq equations in handling flow discontinuities (such as with breaking waves), the model deactivates the dispersion terms during wave breaking and makes use of the underlying NLSW equations where the breaking wave is then approximated as a bore or hydraulic jump. Wave breaking—and the deactivation of the dispersion terms—occurs in the model based on the momentum gradient:

$$(h+\eta)\frac{\partial u}{\partial x} > B\sqrt{g(h+\eta)}, \quad (10)$$

where *B* is a calibration coefficient (by default = 0.5). Here, *B* = 0.8 produced the best agreement between model and observations for both cases. This suggests that under these particularly shallow conditions, the wave face becomes very steep prior to breaking. For a detailed overview of the model's sensitivity to this parameter, the reader is pointed to (Roeber, et al., 2010). All other model parameters were kept at their default values.

The grid resolution ($L_0/dx$) was set as 200 for the mild swell-wave case but was reduced to 60 for the steep wind-wave case to ensure model stability. For the steep-wave case, higher grid resolutions and lower *B* values led to instabilities in the form of strong oscillations in surface elevation in the breaking region. This phenomenon, explored extensively by Kazolea and Ricchiuto (2018), is due to the model's hybrid approach to handling wave breaking; that is, where the Boussinesq equations are reduced to the NLSW equations during wave breaking. It should be noted that Boussinesq wave models which take a different (eddy viscosity) approach to wave breaking reportedly show less sensitivity to the grid size (Kazolea and Ricchiuto, 2018); however, this was not evaluated here.

### 2.2.4 XBeach Non-hydrostatic

Like SWASH, XB-NH solves the NLSW equations with a non-hydrostatic pressure correction term (Equations 4 to 7). Here, XBeach version 1.235527 (also known as the "XBeachX" release) is applied in reduced (simplified) two-layer mode, where the non-hydrostatic pressure is assumed constant in the lower (first) layer (De Ridder, 2018). The water depth is divided into two layers with heights $z_1 = \alpha h$ and $z_2 = (1 - \alpha)h$, where $\alpha$ is the layer distribution. The resulting layer-averaged velocities ($u_1$ and $u_2$) are transformed to a depth-averaged velocity ($U$) and a velocity difference ($\Delta u$). Due to the simplified non-hydrostatic pressure in the lower layer, the vertical velocity between layers is neglected. Therefore, only the continuity relation for the upper (second) layer is required:

$$\frac{\partial}{\partial x}[(1 + \alpha)hU + (1 - \alpha)h\alpha\Delta u] + 2w_2 - u_2\frac{\partial \eta}{\partial x} - u_1\frac{\partial z_1}{\partial x} = 0, \qquad (11)$$

To determine the water elevation, the global continuity equation is applied:

$$\frac{\partial \eta}{\partial t} + \frac{\partial hU}{\partial x} = 0, \qquad (12)$$

In order to control the computed location and magnitude of depth-limited wave breaking, a hydrostatic front approximation is applied. With this, the pressure distribution under breaking waves is considered hydrostatic when the local surface steepness exceeds a maximum prescribed value ($\lambda = 0.5$, by default):

$$\frac{\partial \eta}{\partial t} > \lambda, \qquad (13)$$

Here, $\lambda = 0.9$ and $0.7$ produced the best agreement between the model and observations for the mild- and steep-wave cases, respectively. This further supports the statement that for very shallow foreshores, the waves become particularly steep before breaking. All other model parameters were kept at their default values. Additionally, the grid resolution ($L_0/\Delta x$) was set to ~200 and ~180 for the two respective cases.

### 2.2.5 XBeach Surfbeat

XB-SB solves SS-wave motions using the wave-action equation with time-dependent forcing, similar to that of the HISWA model (Holthuijsen, et al., 1989). The model represents the SS-wave frequency spectrum by a single frequency ($f_{rep}$) and the wave-action equation is applied at the timescale of the wave group:

$$\frac{\partial A}{\partial t} + \frac{\partial c_{gx} A}{\partial x} = -\frac{D_w}{\sigma}, \quad (14)$$

$$A(x,t) = \frac{S_w(x,t)}{\sigma(x,t)} \quad (15)$$

$$\sigma = \sqrt{gk \tanh kh} \quad (16)$$

where $A$ is the wave action, $S_w$ is the wave energy density, $\sigma$ is the intrinsic wave frequency, $k$ is the wave number, $D_w$ is a dissipation term to account for wave breaking and $c_{gx}$ is the wave-action propagation speed in the cross-shore direction. To simulate wave breaking, XB-SB applies a dissipation model (Roelvink, 1993), by default, for use with SS-wave groups; and a roller model (Nairn, et al., 1991, Svendsen, 1984) to represent momentum stored in surface rollers which results in a shoreward delay in wave forcing. The radiation stress gradients that result from these variations in wave action exert forces on the water column and drive IG waves and unsteady currents which are solved by the NLSW equations (Equations 4 to 7). Therefore, the model directly simulates wave-driven currents and the run-up and overtopping of IG waves.

$$\bar{D}_w = 2 \frac{\alpha}{T_{rep}} Q_b E_w \frac{H_{rms}}{h}, \quad (17)$$

$$Q_b = 1 - \exp\left(-\left(\frac{H_{rms}}{H_{max}}\right)^{10}\right) \quad (18)$$

where $\bar{D}_w$ is the total (directionally-integrated) wave energy dissipation due to breaking, $T_{rep} = 1/f_{rep}$ is the representative wave period and $Q_b$ is the fraction of breaking waves; the root-mean-square SS-wave height, $H_{rms} = \sqrt{8E_w/\rho g}$; the maximum wave height, $H_{max} = \gamma_r h$; $E_w$ is the wave-group varying SS-wave energy; $\alpha$ is a dissipation (by default = 1) and $\gamma_r$ is the ratio of breaking waves to local water depth (by default = 0.55 but typically used for calibration).

Here, $\gamma_r = 0.45$ and $0.65$ provided the best agreement between the model and observations for the mild swell and steep wind-wave cases, respectively.

XB-SB does not directly produce the SS-wave component of the energy density spectrum, instead it computes the change in SS-wave energy as a change in the bulk $H_{rms}$ parameter, as described above. In order to produce a complete energy density ($C_{\eta\eta}$) spectrum at each gauge location, a JONSWAP distribution was assumed around the peak-frequency ($f_p$), where $\sqrt{8\int_{f_p/2}^{2} C_{\eta\eta}\, df} = H_{rms}$. This SS-wave spectrum (Figure 3b) was then combined with the IG-wave spectrum (Figure 3a)—obtained directly from the computed surface elevation—to produce the complete spectrum (Figure 3c).

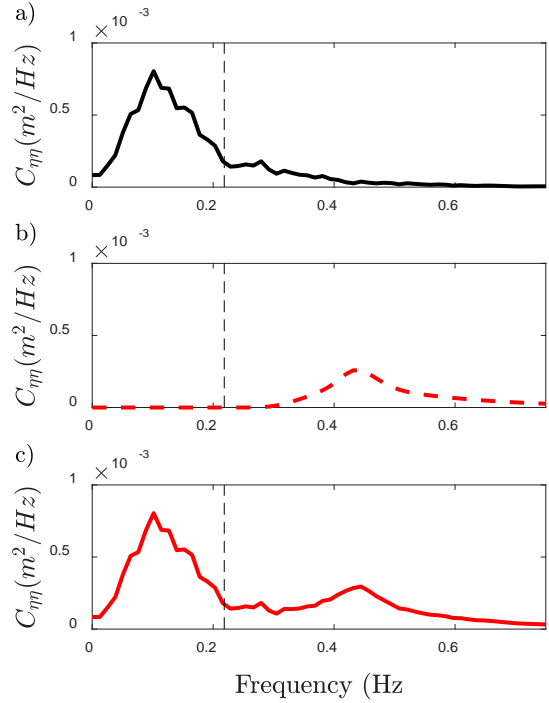

Figure 3 Example of a) the IG-wave spectrum based on the computed surface elevation; b) an assumed SS-wave spectrum (JONSWAP shape) based on the computed root-mean-square SS-wave height ($H_{rms}$); and c) the total combined spectrum, for XB-SB at the dike toe (steep wind-wave case).

For the mild swell-wave case, the grid resolution was varied such that it increased shoreward. This reduced computation time while ensuring that the steep dike slope was accurately capture. As such, $L_0/\Delta x$ varied from ~25 (offshore) to ~160 (at the dike) in the mild-wave case; and from ~45 to ~90 in the steep-wave case.

### 2.2.6 SWAN

SWAN is a third-generation, phase-averaged wave model used to estimate the generation (by wind), propagation and dissipation (by depth-induced breaking and bottom friction) of waves from deep water to the surf zone. This includes wave-wave interactions, in both deep and shallow water, and wave-induced setup; but neglects wave-induced currents and the generation or propagation of IG waves. Like XB-SB, SWAN computes the spectral evolution of $A$ in space and time. This is done in a manner similar to Equation 14; however, unlike XB-SB which makes use of a single representative frequency, SWAN takes the frequency distribution of action density into account. To simulate wave breaking, SWAN uses the following parametric dissipation model (Battjes and Janssen, 1978):

$$\overline{D}_w = \frac{\alpha}{4}\rho g f_{mean} Q_b H_{max}^2, \tag{19}$$

and $Q_b$ is estimated as:

$$\frac{1-Q_b}{\ln Q_b} = -8\frac{E_{tot}}{H_{max}^2}, \tag{20}$$

where $f_{mean}$ is the mean wave frequency, $H_{max} = \gamma_{bj} h$ and $E_{tot}$ is the total wave-energy variance. Here, $\gamma_{bj}$ = 0.73 (default value) provided good agreement between the model and observations for both the mild swell and steep wind-wave cases. For both wave cases, a constant grid spacing of 0.25 m was applied. This corresponded to $L_0/\Delta x \approx 30$ for the mild-wave case and $L_0/\Delta x \approx 20$ for the steep-wave case.

## 2.3 Data Processing and Analysis

### 2.3.1 Mean water level

The mean water level ($\bar{\eta}$) was calculated by taking the average of the surface elevation, $\eta(t)$, at each gauge location, relative to the elevation of the dike toe. The wave-induced setup, <$\eta$>, was then obtained as the difference between $\bar{\eta}$ at each gauge location and $\bar{\eta}$ at the most offshore gauge.

### 2.3.2 Separation of infragravity and sea-swell waves

The time series of $\eta(t)$ were further analysed using the Welch's average periodogram method and a Hann filter with a 50% maximum overlap. The resulting one-dimensional spectra of wave energy density, $C_{\eta\eta}(f)$—with ~43 degrees of freedom and a frequency resolution of ~0.008 Hz—were then used to determine $H_{m0,T}$, $H_{m0,SS}$ and $H_{m0,IG}$, as follows:

$$H_{m0,T} = 4\sqrt{\int_{0.005}^{2} C_{\eta\eta}\, df}, \tag{21}$$

$$H_{m0,SS} = 4\sqrt{\int_{f_p/2}^{2} C_{\eta\eta}\, df}, \tag{22}$$

and

$$H_{m0,IG} = 4\sqrt{\int_{0.005}^{f_p/2} C_{\eta\eta}\, df}, \qquad (23)$$

where half the peak frequency ($f_p/2 = 1/2T_p$) is taken as the cut-off to separate SS and IG motions (Roelvink and Stive, 1989). This choice of cut-off frequency is based on the tendency that, in deep water, the majority of SS-wave energy is found at frequencies $> f_p/2$ while the majority of IG-wave energy lies at frequencies $<f_p/2$.

### 2.3.3 Spectral Wave Period

In addition to wave heights, the spectral wave period ($T_{m-1,0}$) at each gauge location was calculated as follows:

$$T_{m-1,0} = \frac{m_{-1}}{m_0}, \qquad (24)$$

where,

$$m_n = \sqrt{\int_{0.005}^{2} C_{\eta\eta} \cdot f^n\, df}. \qquad (25)$$

### 2.3.4 Empirical Estimate of the Incident Infragravity Waves

As SWAN neglects the contribution of IG waves to the total wave incident height at the dike toe ($H_{m0,T,toe,in}$) we apply an empirical correction, proposed by (Lashley, et al., *Forthcoming*). Using a dataset of 672 XBeach simulations, an empirical formula for the relative magnitude of the IG waves ($\widetilde{H}_{IG}$) was derived with influence factors to account for variations in offshore wave directional spreading ($\bar{\gamma}_\sigma$), $h_{toe}$ ($\bar{\gamma}_h$), $\cot m$ ($\bar{\gamma}_f$), vegetation ($\bar{\gamma}_v$) and $\cot \alpha$ ($\bar{\gamma}_d$):

$$\widetilde{H}_{IG} = 0.36 \cdot H_{m0,T,deep}^{0.5} \cdot \bar{\gamma}_\sigma \cdot \bar{\gamma}_h \cdot \bar{\gamma}_f \cdot \bar{\gamma}_v \cdot \bar{\gamma}_d, \qquad (26)$$

For an incident waves analysis (i.e. without the influence of the dike slope) with no directional spreading (1D flume conditions) or vegetation, $\bar{\gamma}_\sigma, \bar{\gamma}_v$ and $\bar{\gamma}_d = 1$; while,

$$\bar{\gamma}_h = 1.04 \cdot \exp(-1.4 \cdot h_{toe}) + 0.9 \cdot \exp(-0.19 \cdot h_{toe}) \qquad (27)$$

and

$$\bar{\gamma}_f = 1.56 - 3.09 \cdot \cot \alpha_{fore}^{-0.44}. \qquad (28)$$

As $\widetilde{H}_{IG}$ represents the ratio of IG to SS waves, $H_{m0,IG,toe,in}$ can be obtained from a SWAN estimate of $H_{m0,SS,toe,in}$:

$$H_{m0,IG,toe,in} = \widetilde{H}_{IG} \cdot H_{m0,SS,toe,in}. \tag{29}$$

Finally, a corrected estimate of $H_{m0,T,toe,in}$ was obtained as follows:

$$H_{m0,T,toe,in} = \sqrt{H_{m0,SS,toe,in}^2 + H_{m0,IG,toe,in}^2}. \tag{30}$$

### 2.3.5 Empirical Wave Overtopping

While the fully phase-averaged models like SWAN are—to some extent—able to estimate nearshore wave conditions, they cannot directly simulate wave overtopping, as this requires that the individual waves be resolved. In order to estimate wave overtopping, these models can be (and are often) combined with well-established empirical models that require wave parameters at the dike toe as input. In the present study, the EurOtop (2018) formulae based on the work of Van Gent (1999) and Altomare, et al. (2016) for (very) shallow foreshores are applied in combination with SWAN. For smooth dikes under perpendicular wave attack with $h_{toe}/H_{m0,T,deep} < 1.5$:

$$\frac{\bar{q}}{\sqrt{g \cdot H_{m0,T,toe,in}^3}} = 10^{-0.79} \cdot \exp\left(-\frac{R_c}{H_{m0,T,toe,in} \cdot (0.33 + 0.022 \cdot \xi_{m-1,0})}\right), \tag{31}$$

with

$$\xi_{m-1,0} = \frac{\tan\alpha_{sf}}{\sqrt{H_{m0,T,toe,in}/L_{m-1,0}}}, \tag{32}$$

$$L_{m-1,0} = \frac{g \cdot T_{m-1,0,toe,in}^2}{2\pi}, \tag{33}$$

$$\tan\alpha_{sf} = \frac{1.5 H_{m0,T,toe,in} + R_{u2\%}}{(1.5 H_{m0,T,toe,in} - h_{toe}) \cdot m + (h_{toe} + R_{u2\%}) \cdot \cot\alpha}, \tag{34}$$

$$\frac{R_{u2\%}}{H_{m0,T,toe,in}} = 4 - \frac{1.5}{\sqrt{\xi_{m-1,0}}}, \tag{35}$$

where $g$ is the gravitational constant of acceleration, $\alpha_{sf}$ is an equivalent slope (to account for waves breaking on the foreshore) and $T_{m-1,0,toe,in}$ is the spectral wave period at the dike toe based on the incident waves (i.e. without the influence of waves reflected at the dike). It should be noted that $\xi_{m-1,0}$ and $R_{u2\%}$ are obtained iteratively (until $R_{u2\%}$ converges), with a first estimate of $R_{u2\%} = 1.5 H_{m0,T,toe,in}$.

Additionally, as SWAN excludes the contribution of IG waves, corrected estimates of $T_{m-1,0,toe}$ are typically obtained using Equations 36 and 37 (Hofland, et al., 2017), as outlined in the EurOtop (2018) manual:

$$\frac{T_{m-1,0,toe,in}}{T_{m-1,0,deep}} - 1 = 6 \cdot \exp(-4\tilde{h}) + \exp(-\tilde{h}), \tag{36}$$

where,

$$\tilde{h} = \frac{h_{toe}}{H_{m0,T,deep}} \left(\frac{\cot m}{100}\right)^{0.2}. \tag{37}$$

### 2.3.6 Error Metrics

In order to compare the performance of the numerical models, we assess the mean relative accuracy in an approach similar to that of Lynett, et al. (2017):

$$Mean\ Ratio_\Psi = \frac{1}{n}\sum_{i=1}^{N} \frac{\Psi_{mod}^i}{\Psi_{obs}^i}, \tag{38}$$

where $\Psi$ is a stand-in for the parameter under consideration ($\bar{\eta}$, $H_{m0,T}$, $H_{m0,SS}$, $H_{m0,IG}$ and $T_{m-1,0}$) for the $N$ wave-gauge locations; and subscripts $mod$ and $obs$ refer to model predictions and observations made during the physical experiment, respectively. Here we make a distinction between gauges offshore and nearshore (Figure 2) A mean ratio of 1 suggests perfect agreement between the model and observations, while values higher or lower than one indicate over- or under-predictions, respectively. It should be noted that all wave gauges (offshore and at the dike toe, see Figure 2) are considered in Equation 38. While the focus of this study is primarily at the dike toe, it is important to assess the model performance offshore to ensure that: i) the boundary conditions are correctly modelled; and ii) that no (significant) numerical dissipation occurs in deep water, as a result of a coarse grid resolution for example.

Finally, the performance of each model for wave overtopping was also assessed by comparing the absolute relative error in the prediction of mean overtopping discharge:

$$Absolute\ Relative\ Error_{\bar{q}} = \left|1 - \frac{\bar{q}_{mod}}{\bar{q}_{obs}}\right|. \tag{39}$$

### 2.3.7 Computation Speed

Two work stations (WS) were used to carry out this research (Table 3). Given the required computational effort, the OpenFOAM simulations were performed on WS-A, while the other models were run on WS-B. To assess computation speed, the duration of each simulation (in wall clock time) was recorded.

Table 3 Overview of work stations used to carry out the numerical simulations.

| Work Station (WS) | A | B |
|---|---|---|
| Operating System | Ubuntu 14.04 LTS | Windows 10 |
| Memory | 31.2 GB | 16 GB |
| Processor | Intel Xeon® CPU ES-2690 v3 @ 2.60 GHz x 16 | Intel® Core™ i7-6600 CPU @ 2.60GHz, 2.81 GHz x 4 |
| Graphics | Gallium 0.4 on NVE7 | Intel® HD Graphics 520 |
| Type | 64-bit | 64-bit |
| Disk | 1.9 TB | 239 GB |

## 3 Results and Discussions

In this section, the results of the model-data comparisons are presented and discussed. As wave overtopping is the end result of wave propagation, the performance of each model for the prediction of mean water levels, wave heights and periods is first assessed. For the models where calibration was carried out (BOSZ, XB-NH and XB-SB), both default and calibrated results are presented. Note that no parameter tuning was done for the depth-resolving models (OpenFOAM and SWASH) as wave breaking is intrinsically resolved. Likewise, SWAN with default settings showed reasonable agreement and was therefore not calibrated. Lastly, it should be noted that the BOSZ simulation of the steep wind-wave case with default settings resulted in instabilities (see Section 2.2.3) and is therefore not included in the analysis.

### 3.1 Mean Water Level

Each model, excluding OpenFOAM, is able to accurately (within 15% error) and consistently reproduce $\bar{\eta}$ for both the mild swell (Figure 4a) and steep wind-wave (Figure 4b) cases. This includes the increase in $\bar{\eta}$ nearshore, referred to as wave-induced setup ($<\eta>$), highlighted in Figure 5 with the XB-NH results representing the general behaviour of the numerical models. While OpenFOAM agrees well with the observations for the mild swell case, it overestimates $<\eta>$ offshore (gauges 2 and 3) and underestimates $<\eta>$ nearshore (gauges 4 to 6) for the steep wind-wave case (Figure 5). This may be indicative of premature wave breaking in OpenFOAM.

The satisfactory performance of BOSZ and XB-NH observed here (Figure 4) is in contrast with previous studies (Lashley, et al., 2018, Zhang, et al., 2019), which found that depth-averaged models were unable to accurately estimate <$\eta$> due to their lack of vertical resolution and exclusion of wave roller dynamics. However, the difference in model performance here is likely due to the spilling nature of the waves ($\xi_{0,fore}$ = 0.23 and 0.13 for the mild swell and steep wind-wave cases, respectively) compared to the plunging waves and steep fore-reef slopes assessed by Lashley, et al. (2018) ($\xi_{0,fore}$ > 1.1) and Zhang, et al. (2019) ($\xi_{0,fore}$ = 1.29). That is, while resolving the vertical structure of flow may be critical for plunging breakers, depth-averaged models are able to perform well under spilling waves.

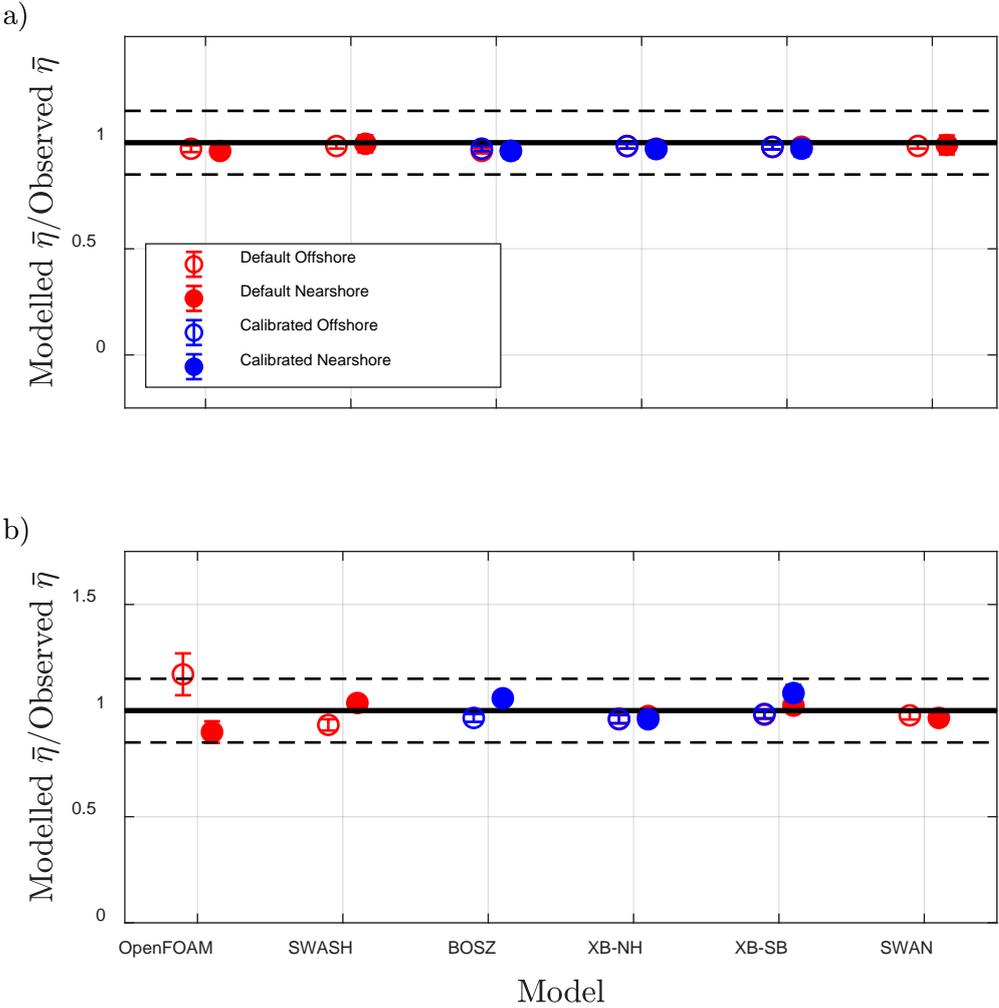

Figure 4 Mean ratio of modelled to observed $\bar{\eta}$ (markers) for both the a) mild swell and b) steep wind-wave cases, with error bars representing the standard deviation. Solid horizontal lines represent perfect agreement between model and observations. Dashed lines correspond to +/- 15% error.

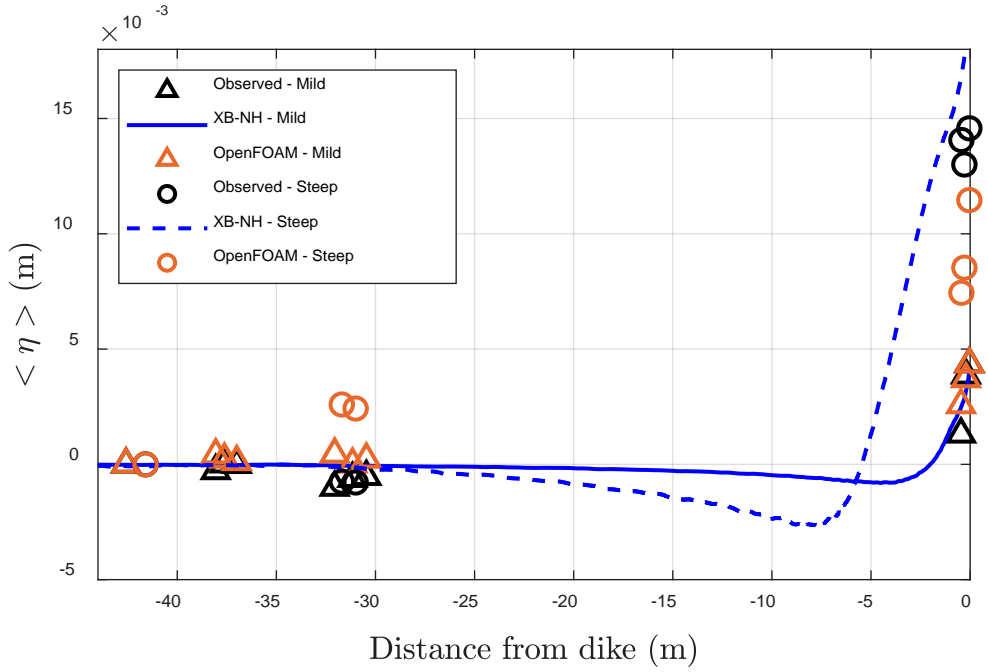

Figure 5 Cross-shore profiles of modelled (XB-NH and OpenFOAM) and observed <η> for both the mild swell and steep wind-wave cases.

There is also a notable difference in the observed maximum <η> between the mild swell (<η> = 0.004 m) and steep wind-wave (<η> = 0.015 m) cases (Figure 5). This substantial increase in <η> as $h_{toe}/H_{m0,T,deep}$ decreases agrees with the findings of Gourlay (1996) on shallow reefs, and suggests that <η>—which contributes to wave run-up (Stockdon, et al., 2006) and, by extension, overtopping—increases proportionally as foreshores become more shallow, or as deep water wave conditions become more energetic.

## 3.2 Significant Wave Height

SWASH, BOSZ, XB-NH and XB-SB are able to reproduce $H_{m0,T}$, both offshore and nearshore, within 15% error for two cases (Figure 6). On the other hand, OpenFOAM and SWAN both show notable differences; with SWAN consistently and considerably underestimating $H_{m0,T}$ nearshore.

While SWAN is able to accurately simulate the propagation of high-frequency waves ($H_{m0,SS}$, Figure 7), it does not compute the low-frequency waves ($H_{m0,IG}$, Figure 8) and therefore underestimates $H_{m0,T}$ nearshore, where the contribution of IG waves is significant (Figure 6). The relatively high standard deviation associated with SWAN's nearshore $H_{m0,SS}$ estimates is due to its exclusion of wave reflection. In the physical model, the superposition of the incident and reflected waves results in a nodal/anti-nodal pattern with a maximum at the dike (outsets in Figure 9a and Figure 10a). As SWAN excludes the reflected component, the model underestimates $H_{m0,SS}$ immediately in front of the dike, where the incident and reflected waves add up. On the other hand, this shortcoming makes SWAN especially suitable for use with the empirical overtopping models that require incident-wave conditions as input.

SWAN also predicts a higher and lower maxima in $H_{m0,SS}$ (just before breaking) than XB-NH, for the mild swell (Figure 9a) and steep wind-wave (Figure 10a) cases, respectively. This is likely due to the dissipation model employed by SWAN (Equation 19). Tuning $\gamma_{bj}$—the parameter which controls the maximum wave height to water depth ratio in SWAN—would yield better agreement between the two models; however as there were no wave gauges in this region it is difficult to ascertain which model is correct here.

With respect to OpenFOAM, the model shows inconsistent results between the two cases. Under the mild swell conditions, OpenFOAM underestimates $H_{m0,T}$ nearshore (Figure 6a); however for the steep wind-wave case, the model overestimates $H_{m0,T}$ nearshore (Figure 6b). In both cases, the model appears to be too dissipative, resulting in a reduction in $H_{m0,SS}$ offshore. For the mild swell case, this dissipation is minor resulting in a consistent under-prediction of $H_{m0,SS}$ (Figure 7a and Figure 9a) and $H_{m0,IG}$ (Figure 8a and Figure 9b). Under the steep wind-wave conditions, however, the dissipation is significant. This observation, combined with the overestimation of $<\eta>$ offshore (Figure 5), indicates premature wave breaking in OpenFOAM. This reduction in $H_{m0,SS}$ offshore results in unbroken SS-waves reaching the dike and the overestimation of $H_{m0,SS}$ nearshore (Figure 7b and Figure 10a). XB-NH also shows some numerical dissipation offshore but this is negligible compared to that of OpenFOAM (Figure 10a). As a reduction in grid size did not significantly improve the OpenFOAM model results, the observed dissipation is possibly due to an over-production of turbulence leading to premature wave decay (Larsen and Fuhrman, 2018).

Though SWASH is able to accurately predict $H_{m0,IG}$ it underestimated $H_{m0,SS}$ nearshore in both cases. This is possibly due to the standard k-ε turbulence model applied in multi-layered mode, which may overestimate the turbulent (vertical) viscosity. Similar to OpenFOAM, a reduction in grid size from 0.04 m to 0.025 m did not significantly improve the estimates (~3% change in $H_{m0,SS}$). While calibration generally improved model performance for $\bar{\eta}$ (Figure 4), $H_{m0,T}$ (Figure 6) and $H_{m0,IG}$ (Figure 8), it resulted in the overestimation of $H_{m0,SS}$ nearshore by XB-SB (Figure 7). This is as a result of tuning $\gamma_r$ (Equation 18) which affects both the maximum $H_{m0,SS}$ and $H_{m0,IG}$. Perhaps a different approach, where $\alpha$ (Equation 17)—the parameter that controls the magnitude of dissipation—is calibrated (Lashley, et al., 2018) would yield better results. However, as XB-SB predicts IG-wave overtopping only, the loss in accuracy for $H_{m0,SS}$ to improve $H_{m0,IG}$ predictions was considered acceptable.

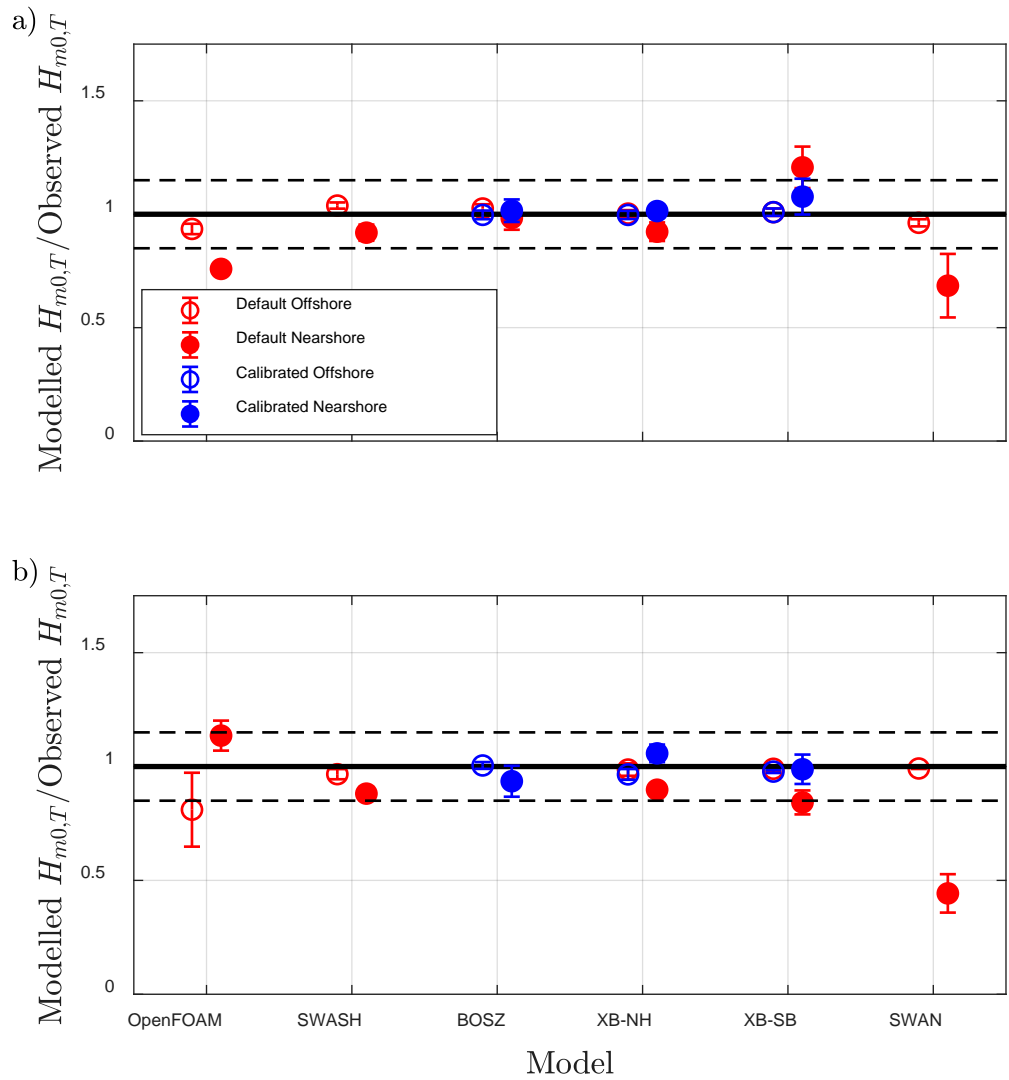

Figure 6 Mean ratio of modelled to observed $H_{m0,T}$ (markers) for both the a) mild swell and b) steep wind-wave cases, with error bars representing the standard deviation. Solid horizontal lines represent perfect agreement between model and observations. Dashed lines correspond to +/- 15% error.

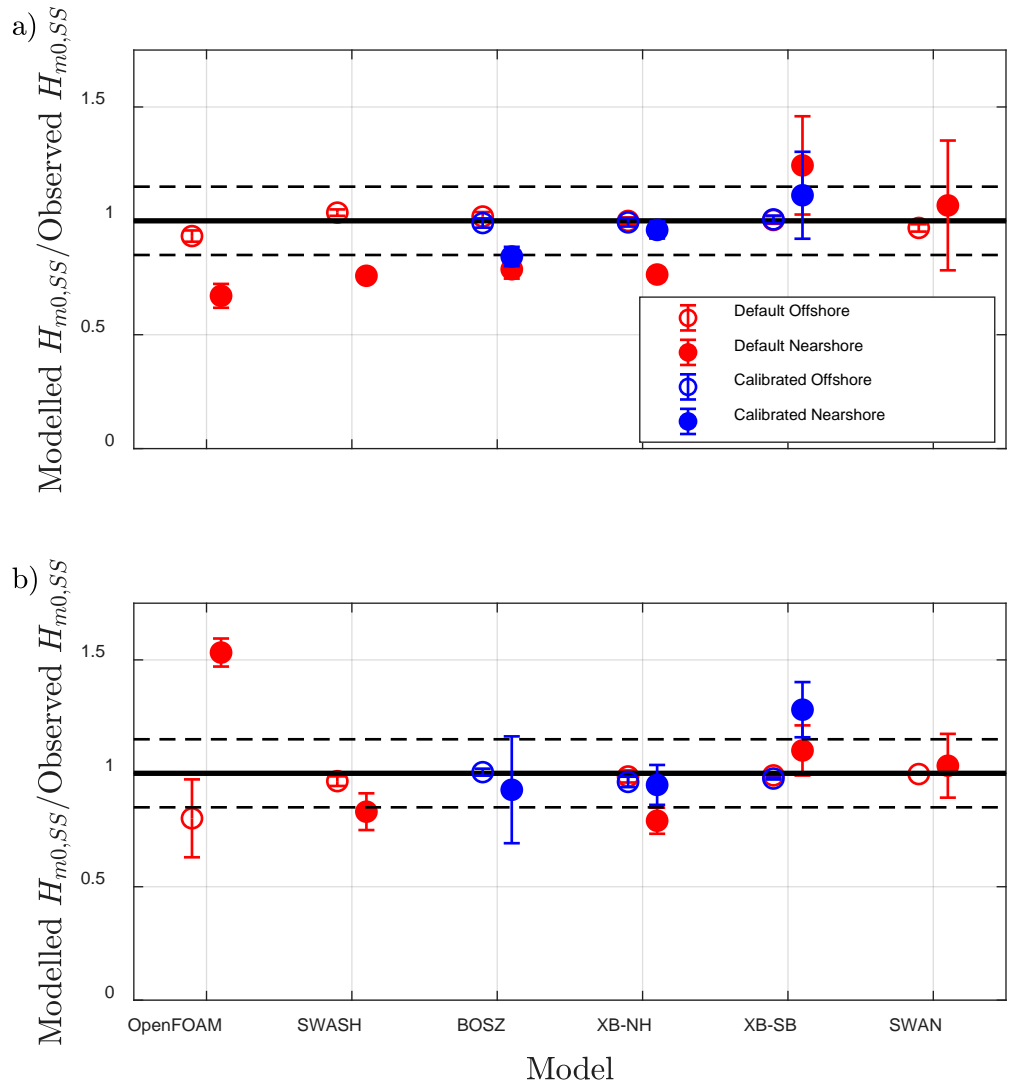

Figure 7 Mean ratio of modelled to observed $H_{m0,SS}$ (markers) for both the a) mild swell and b) steep wind-wave cases, with error bars representing the standard deviation. Solid horizontal lines represent perfect agreement between model and observations. Dashed lines correspond to +/- 15% error.

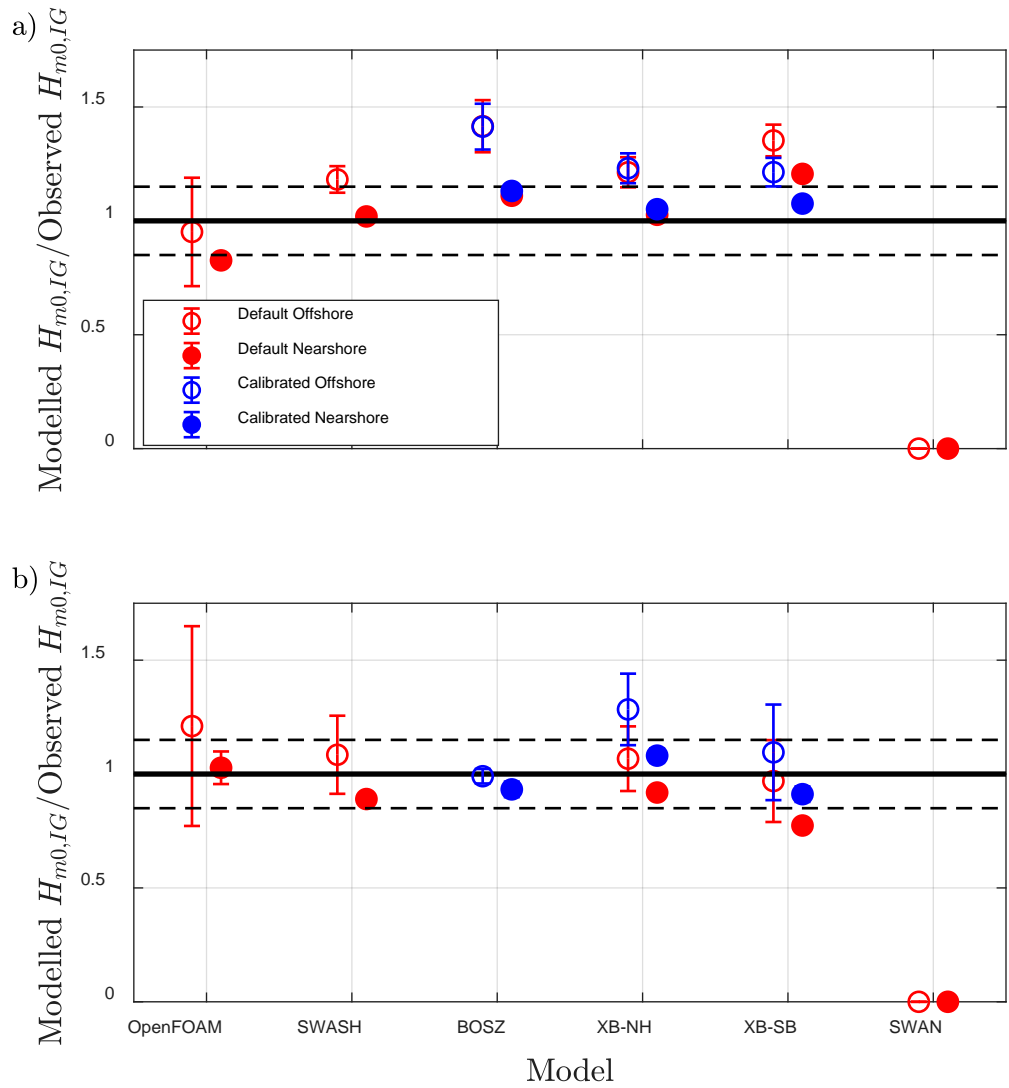

**Figure 8** Mean ratio of modelled to observed $H_{m0,IG}$ (markers) for both the a) mild- and b) steep-wave cases, with error bars representing the standard deviation. Solid horizontal lines represent perfect agreement between model and observations. Dashed lines correspond to +/- 15% error.

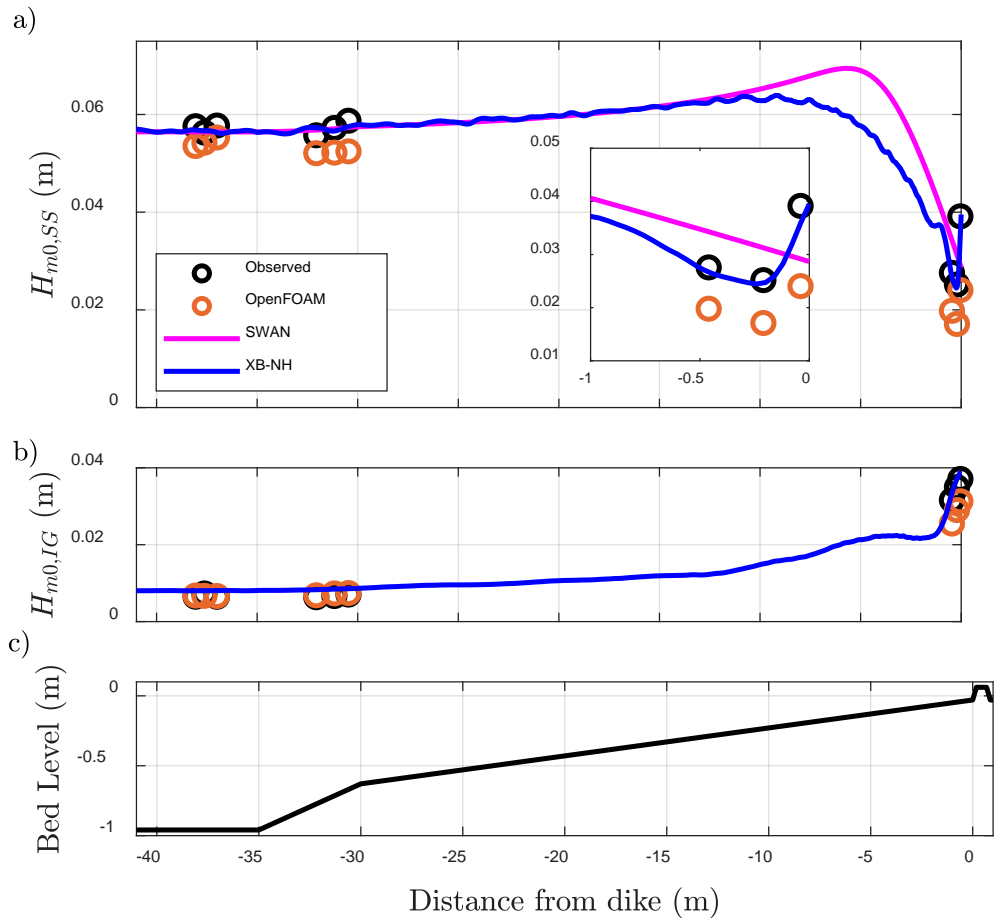

Figure 9 Cross-shore profiles of modelled and observed: a) $H_{m0,SS}$ and b) $H_{m0,IG}$ for the mild-swell case; with c) bed level, for reference. Outset in panel 'a' magnifies the plot area between -1 and 0 m away from the dike.

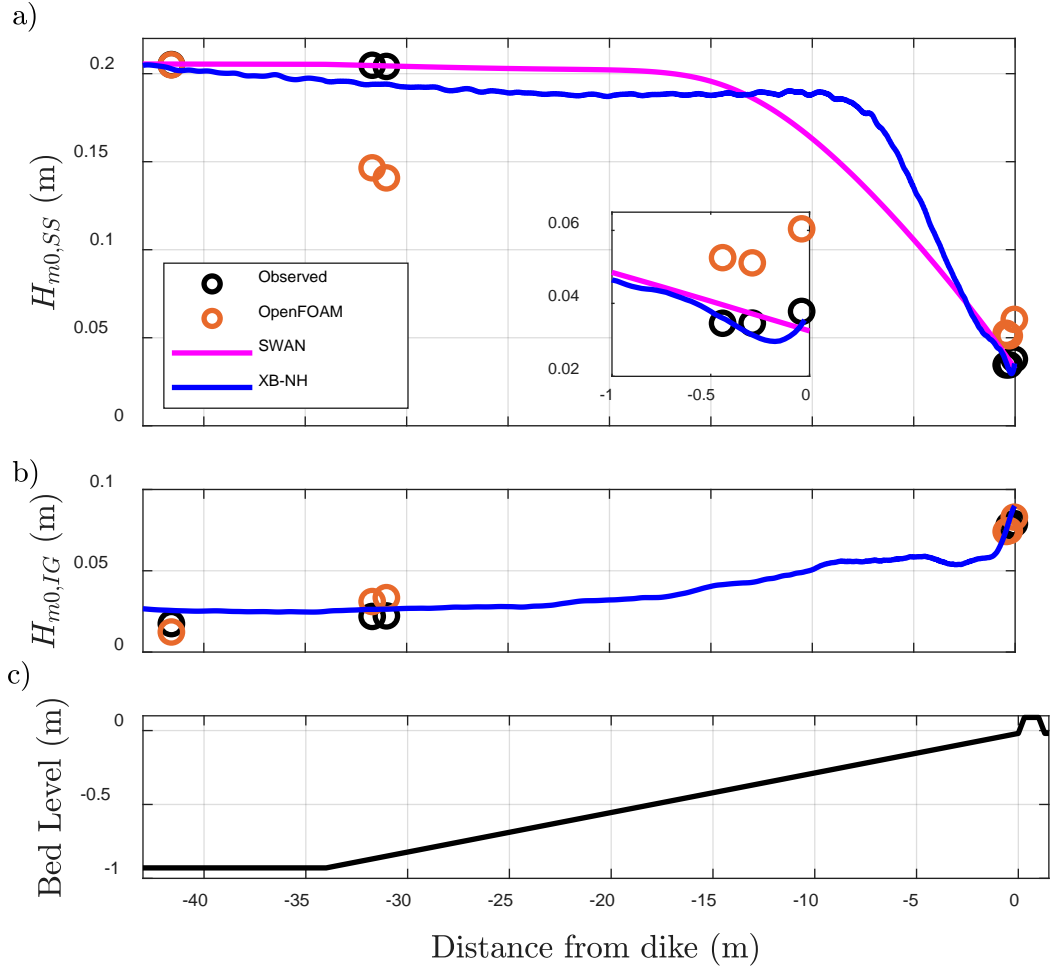

Figure 10 Cross-shore profiles of modelled and observed: a) $H_{m0,SS}$ and b) $H_{m0,IG}$ for the steep wind-wave case; with c) bed level, for reference. Outset in panel 'a' magnifies the plot area between -1 and 0 m away from the dike.

### 3.3 Spectral Wave Period

SWASH, XB-NH and XB-SB show good agreement between modelled and observed $T_{m-1,0}$ predictions; while SWAN, OpenFOAM and BOSZ show notable deviations. As the accurate prediction of $T_{m-1,0}$ requires the models to correctly represent the distribution of wave energy by frequency (Equation 25), we assess the modelled versus observed wave spectra (Figure 12). SWASH, BOSZ, XB-NH and XB-SB correctly capture the shift in peak energy density ($C_{\eta\eta}$) from the SS-wave (Figure 12a and b) to the IG-wave band (Figure 12c and d); however BOSZ overestimates the magnitude of the IG peak and shows it at slightly lower frequencies than observed. This, coupled with a minor underestimation of the SS-wave energy—most evident for the mild swell case (Figure 12c)—results in an overestimation of $T_{m-1,0}$.

The consistent underestimation of $T_{m-1,0}$ by SWAN is expected due to its exclusion of $C_{\eta\eta}$ at IG frequencies (Figure 12c and d). OpenFOAM, on the other hand, does show a shift in energy from offshore to the dike toe; however, it shows two distinct IG peaks (Figure 12c and d), not present in the observations. In the mild swell case, this misrepresentation of $C_{\eta\eta}$ at IG frequencies couple with the underestimation of $C_{\eta\eta}$ in the SS-wave band resulted in the significant overestimation of $T_{m-1,0}$ nearshore (Figure 11a). Under the steep wind-wave conditions, OpenFOAM also shows considerable $C_{\eta\eta}$ in the SS-wave band (nearshore) while the observed spectra shows very little (Figure 12d). This further supports the argument that due to premature wave decay in the model, some unbroken SS waves are able to reach the dike.

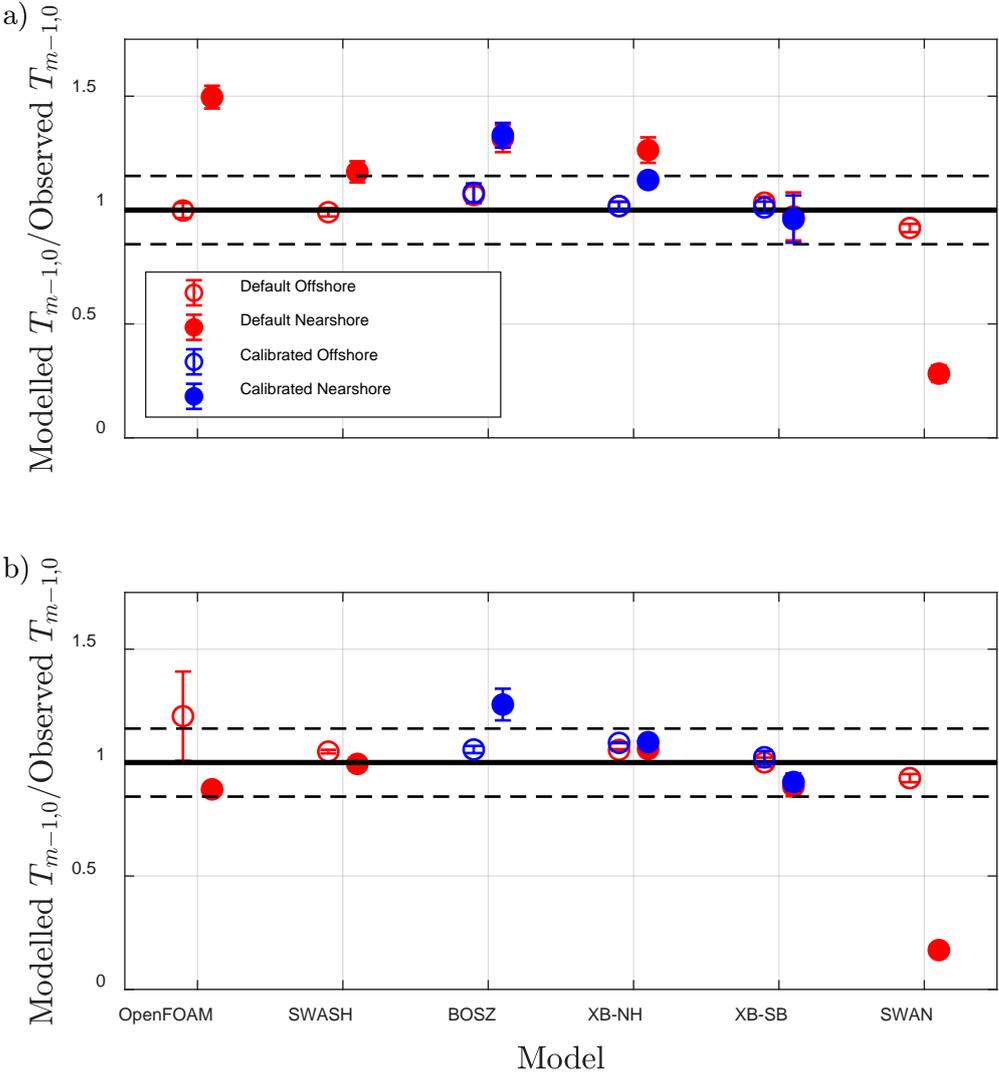

Figure 11 Mean ratio of modelled to observed $T_{m-1,0}$ (markers) for both the a) mild swell and b) steep wind-wave cases, with error bars representing the standard deviation. Solid horizontal lines represent perfect agreement between model and observations. Dashed lines correspond to +/- 15% error.

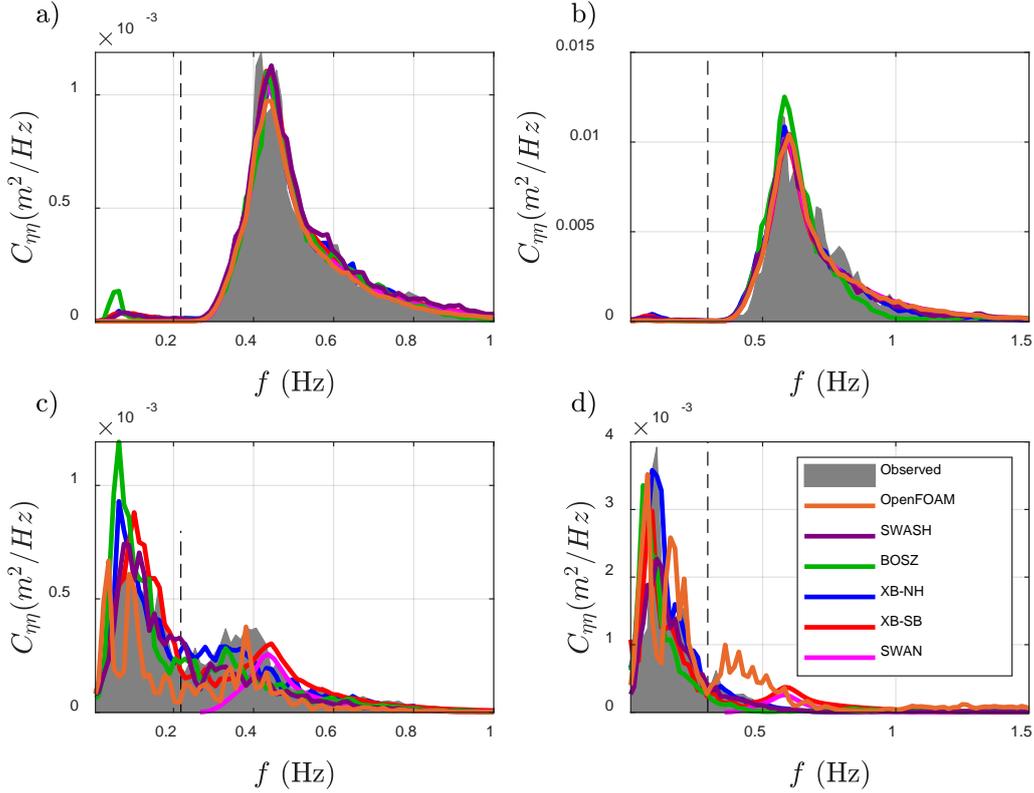

Figure 12 Model-data comparison of wave spectra: a) offshore (at gauge 1) and c) nearshore (at gauge 10) for the mild swell case; and b) offshore (at gauge 1) and d) nearshore (at gauge 6) for the steep wind-wave case. Dashed vertical lines indicate the frequency separating SS- and IG-wave motions.

## 3.4 Wave Overtopping

### 3.4.1 General

An important remark is the difference in the observed $\bar{q}$ between the mild swell (0.094 l/s per m) and steep wind-wave (0.205 l/s per m) cases. Despite having similar $H_{m0,SS,toe}$ values—0.039 m (Figure 9) and 0.038 m (Figure 10a) for the mild swell and steep wind-wave cases, respectively—the steep-wave case with a higher $R_c$ and lower $h_{toe}$ (Table 2), produces double the $\bar{q}$. This observation suggests that the nonlinear effects of wave breaking—that is, the generation of IG-waves and wave-induced setup—contribute significantly to the resulting overtopping discharge. While the effects of vegetation are not considered here, this observation highlights a potential limitation in studies that assess the effectiveness of shallow foreshores but focus only on the attenuation of $H_{m0,SS}$ and neglect the contribution of $H_{m0,IG}$ (Vuik, et al., 2016, Yang, et al., 2012).

To further investigate the influence of the IG-waves, we compare the overtopping estimated using SWAN and EurOtop with and without the corrections to $H_{m0,IG,toe,in}$ and $T_{m-1,0,toe,in}$ obtained through Equations 29 and 36, respectively:

Table 4 SWAN results with and without the empirial corrections for $H_{m0,IG,toe,in}$ (Equation 29) and $T_{m-1,0,toe,in}$ (Equation 36).

| Case | SWAN | $H_{m0,SS,toe,in}$ (m) | $H_{m0,IG,toe,in}$ (m) | $H_{m0,T,toe,in}$ (m) | $T_{m-1,0,toe,in}$ (s) | $\bar{q}$ (l/s per m) Modelled | $\bar{q}$ (l/s per m) Observed |
|---|---|---|---|---|---|---|---|
| Mild swell | Original | 0.029 | 0 | 0.029 | 2.17 | 0.014 | 0.094 |
| | Corrected | 0.029 | 0.014 | 0.032 | 5.38 | 0.053 | |
| Steep wind wave | Original | 0.033 | 0 | 0.033 | 1.67 | 0.003 | 0.205 |
| | Corrected | 0.033 | 0.03 | 0.045 | 8.37 | 0.089 | |

Under the mild swell-wave conditions, the ratio $\widetilde{H}_{IG} = 0.5$ (Equation 29) and the contribution of $H_{m0,IG,toe,in}$ to $H_{m0,T,toe,in}$ is minor (Table 4). On the other hand, including the IG waves resulted in a 2.5-fold increase in $T_{m-1,0,toe,in}$ and magnitude 4-fold increase in the predicted $\bar{q}$, compared to the original SWAN estimates. The difference is more striking for the steep-wave case where the IG waves dominate at the dike toe ($\widetilde{H}_{IG} = 0.92$). The inclusion of the IG waves resulted in 36% increase in $H_{m0,T,toe,in}$, a 5-fold increase in $T_{m-1,0,toe,in}$ and an order of magnitude increase in the predicted $\bar{q}$. Furthermore, the original SWAN estimates—without any corrections to $T_{m-1,0,toe,in}$ and $H_{m0,T,toe,in}$—erroneously show a decrease in $\bar{q}$ between the mild swell and steep wind-wave cases, while the observations show a notable increase. These results further emphasize the danger of neglecting the IG-wave contribution—demonstrated here by the correction of input to the empirical formulae—in the design and assessment of coastal structures with very shallow foreshores.

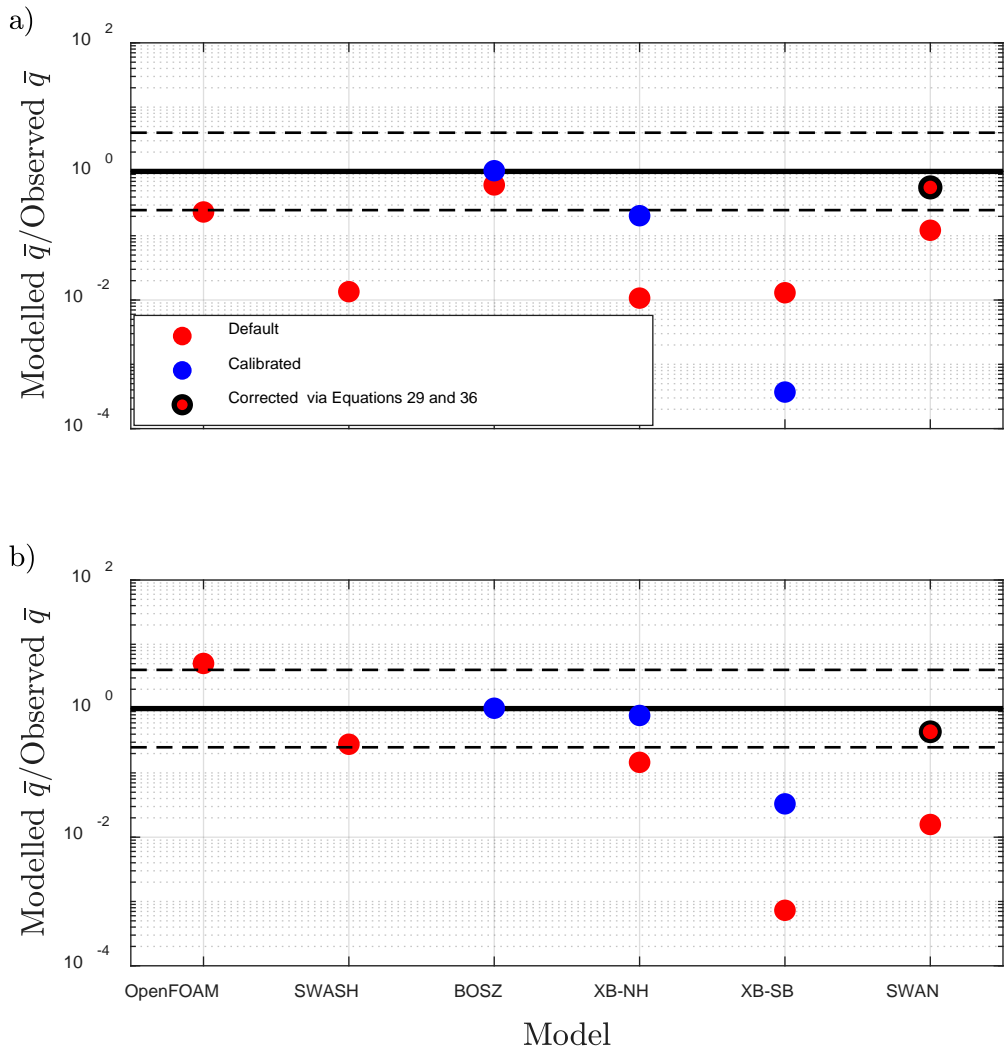

**Figure 13** Mean ratio of modelled to observed $\bar{q}$ (markers) for both the a) mild swell and b) steep wind-wave cases, with error bars representing the standard deviation. Solid horizontal lines represent perfect agreement between model and observations. Dashed lines correspond to a factor of 4 larger and lower than the observations.

Considering the wider model comparison, each model—with the exception of BOSZ—fails to reproduce the overtopping for the mild swell case. This is particularly evident for SWASH and XB-SB which significantly underestimate $\bar{q}$ for both wave cases, with the calibrated XB-SB model producing zero overtopping. This suggests that while XB-SB may estimate wave run-up accurately in IG-wave dominant environments (Lashley, et al., 2018), it's exclusion of the SS-wave component considerably limits its performance for wave overtopping.

The poor performance of SWASH here for wave overtopping is surprising, since it performed reasonably well in the prediction of $\bar{\eta}$, $H_{m0}$ and $T_{m-1,0}$ in both cases here and has been previously successful in one-layered mode (Suzuki, et al., 2017). However, Suzuki, et al. (2017) focused on obtaining good agreement at the toe (the last wave gauge only) and the resulting $\bar{q}$ in their tuning of SWASH; therefore $\bar{q}$ was not assessed unless wave heights and periods at the toe were within a certain accuracy range, regardless of the input (offshore) conditions. Whereas here, we assess the model's general performance for wave propagation (both offshore and nearshore), in addition to $\bar{q}$. It should be noted that a finer grid resolution had little impact on SWASH predictions of $\bar{\eta}$, $H_{m0}$ and $T_{m-1,0}$ (~3%), it increased $\bar{q}$ by a factor of 7—though still significantly underestimated (not shown)—for the mild swell case, with significantly increased computational demand. The models do, however, perform considerably better for the steep-wave case. This is consistent with the findings of Roelvink, et al. (2018) and Suzuki, et al. (2017) who showed that XB-NH and SWASH, respectively, were more accurate for higher overtopping rates, but suffered for rates below 0.08 – 0.16 l/s per m (in model scale).

The improvement in SWAN with the corrections is most evident for the steep-wave case, with the estimated $\bar{q}$ now on par with that of BOSZ and outperforming the other more physically-complex models. Figure 14 shows the modelled relative overtopping discharge ($\bar{q}/\sqrt{gH_{m0,T,toe,in}^3}$) versus the relative freeboard ($R_c/(H_{m0,T,toe,in}(0.33+0.022\xi_{m-1,0,sf}))$) where $H_{m0,T,toe,in}$ and $T_{m-1,0}$ (to compute $\xi_{m-1,0,sf}$) are taken from Table 4. If we take the +/5 % exceedance lines of the EurOtop formula (Equation 31, Figure 14) as the general range of acceptable overtopping predictions, OpenFOAM, BOSZ, XB-NH and SWAN (with corrections) are all reasonable. SWASH and XB-SB, on the other hand, underestimate $\bar{q}$ and fall outside this acceptable range.

As most of the models performed reasonably well for wave propagation, the excellent agreement between BOSZ and the observed $\bar{q}$ is likely not dependent on underlying governing equations (Boussinesq versus NLSW) but more to do with how the shoreline and wave run-up are treated numerically. However, an in-depth analysis of the various numerical schemes implemented in each numerical model was beyond the scope of this study.

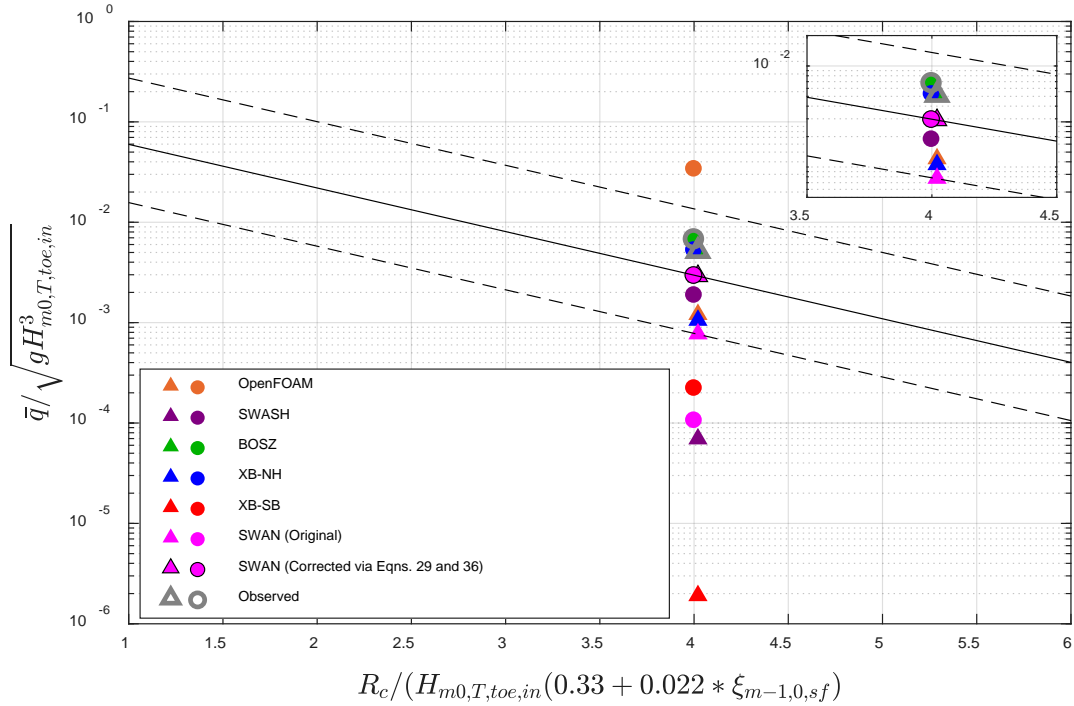

Figure 14 Relative overtopping discharge versus relative freeboard. Solid line corresponds to Equation 31 with dashed lines representing +/- 5% exceedance. Outset magnifies plot area between $\bar{q}/\sqrt{gH^3_{m0,T,toe,in}}$ = 5 x 10$^{-3}$ and 2 x 10$^{-2}$.

### 3.4.2 Accuracy versus Speed

In contrast with the general assumption that models of increasing physical complexity produce more accurate results, Figure 15 shows no clear relationship between computational demand (simulation time) and the absolute relative error in overtopping. Furthermore, the depth-resolving models (SWASH and OpenFOAM), which have significantly higher simulation times show larger errors than the depth-averaged models (XB-NH and BOSZ). The phase-averaged models (XB-SB and SWAN (original)), despite their considerable speed advantage, significantly underestimated the overtopping discharge due to their exclusion of higher- and lower-frequency wave components, respectively. However, by including the IG-waves empirically, SWAN's performance improved significantly; now within acceptable limits and on par with those of XB-NH and BOSZ but at little to no computational cost (Figure 15). It should be noted that the use of SWAN with Equation 36 is already the recommended approach in EurOtop (2018); the novelty here is the further improvement in results offered by Equation 29.

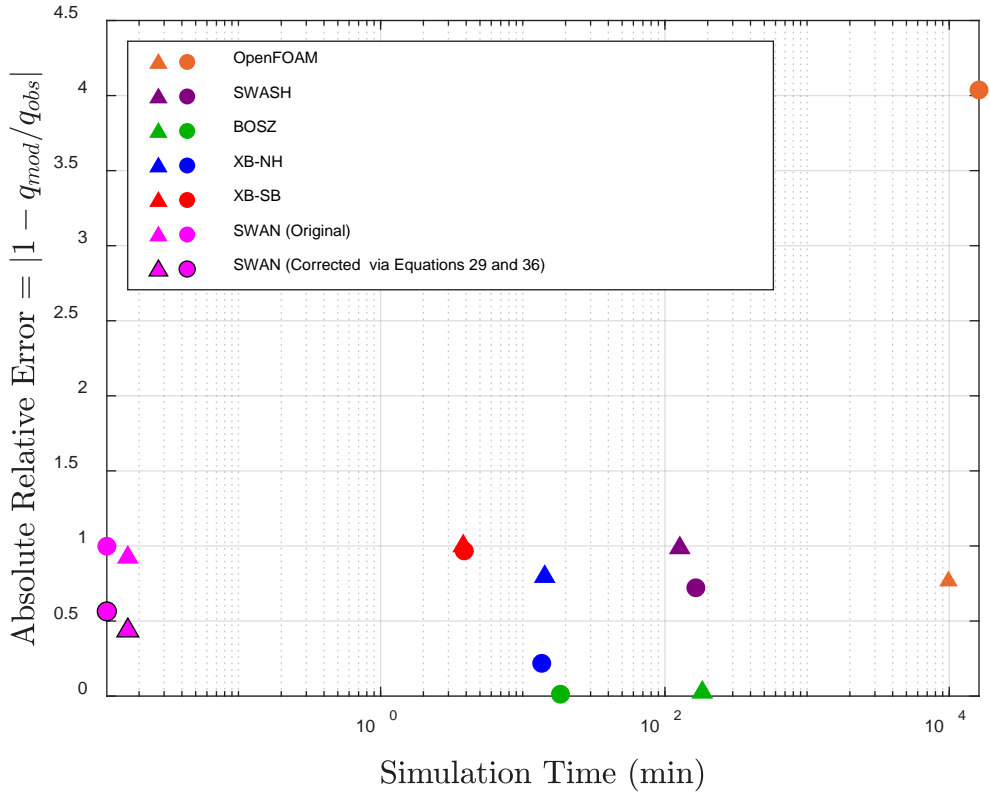

Figure 15 Accuracy versus speed of the numerical wave models for wave overtopping with triangles and circles representing the mild swell and steep wind-wave cases, respectively.

## 4 Conclusion

In the present study we assess the ability of 6 widely used numerical models to simulate waves overtopping steep dikes with mildly-sloping shallow foreshores. However, with the exception of OpenFOAM and to some extent SWASH (multi-layered mode) the above (phase-resolving) models were originally developed to simulate wave evolution over mildly-sloping foreshores; and not specifically for wave run-up and overtopping of steep structure slopes. Since their development, the phase-resolving models have each been successfully applied to wave propagation over steep reefs and run-up of relatively steep beaches. Likewise, depth-resolving models like OpenFOAM and SWASH (multi-layered) were originally developed to simulate wave-structure interaction and not specifically for wave propagation. In the present study we test the ability of these models in both applications: i) wave evolution over a shallow mildly-sloping foreshore; and ii) the resulting overtopping discharge.

Overall, BOSZ and XB-NH (under steep wind-waves) showed high skill in both applications with a reasonable computational demand; while OpenFOAM, with a much higher computational demand—showed difficulty in performing both functions. The broad implication of the present work is that higher-resolution, more computationally-demanding wave models may simply not be needed; specifically where the analysis is focused on bulk, time-averaged physical quantities ($H_{m0}$, $T_{m-1,0}$ and $\bar{q}$), as shown here. Should more detail be required—for example, estimates of the vertical velocity profile or turbulence—then a depth-resolving model such as SWASH (multi-layer) or OpenFOAM should be applied. Moreover, SWASH and OpenFOAM are likely to perform well if the computational domain begins at the dike toe and ends at the overtopping box; i.e., where simulating wave propagation over a large domain is not required.

In addition, our results showed that with simple empirical corrections, phase-averaged models like SWAN can perform on par—if not better than—phase-resolving models, with much less computational effort. Importantly, our work emphasizes the importance of including IG waves in the design and assessment of coastal dikes; as neglecting their contribution to $H_{m0,T,toe}$ and $T_{m-1,0,toe}$ can lead to under-predictions in $\bar{q}$ of up to two orders of magnitude.

Given the scope of the model comparison, including both phase-resolving and phase-averaged, a detailed wave-by-wave comparison of the higher-resolution models was not carried out. Future work should address this and investigate the influence of the various numerical schemes implemented in the respective numerical models, as this was not within the scope of the present work. Additionally, Equations 31 and 36 were developed (in part) using the wider dataset from which these cases were taken; therefore their performance under different conditions is still to be confirmed. Despite these limitations, the findings here can aid practitioners in their decision making; specifically in deciding which numerical model should be applied based on the level of accuracy required.

## Acknowledgements

This work is part of the *Perspectief* research programme *All-Risk* with project number B2 which is (partly) financed by NWO Domain Applied and Engineering Sciences, in collaboration with the following private and public partners: the Dutch Ministry of Infrastructure and Water Management (RWS); Deltares; STOWA; the regional water authority, Noorderzijlvest; the regional water authority, Vechtstromen; It Fryske Gea; HKV consultants; Natuurmonumenten; and waterboard HHNK. Dr. Corrado Altomare acknowledges funding from the European Union's Horizon 2020 research and innovation programme under the Marie Sklodowska-Curie grant agreement No.: 792370. Volker Roeber acknowledges financial support from the Isite program Energy Environment Solutions (E2S), the Communauté d'Agglomération Pays Basque (CAPB) and the Communauté Région Nouvelle Aquitaine (CRNA) for the chair position HPCWaves and the European Union's Horizon 2020 Research and Innovation Programme. The authors also acknowledge Akshay Patil for his incite and efforts towards the success of the research.

# References

Akbar, M., and Aliabadi, S. (2013). "Hybrid numerical methods to solve shallow water equations for hurricane induced storm surge modeling." *Environmental Modelling & Software*, 46, 118-128.

Altomare, C., Domínguez, J. M., Crespo, A. J. C., Suzuki, T., Caceres, I., and Gómez-Gesteira, M. (2015). "Hybridization of the Wave Propagation Model SWASH and the Meshfree Particle Method SPH for Real Coastal Applications." *Coastal Engineering Journal*, 57(04), 1550024.

Altomare, C., Suzuki, T., Chen, X., Verwaest, T., and Kortenhaus, A. (2016). "Wave overtopping of sea dikes with very shallow foreshores." *Coastal Engineering*, 116, 236-257.

Altomare, C., Tagliafierro, B., Dominguez, J. M., Suzuki, T., and Viccione, G. (2018). "Improved relaxation zone method in SPH-based model for coastal engineering applications." *Applied Ocean Research*, 81, 15-33.

Battjes, J. A., and Janssen, J. (1978). "Energy loss and set-up due to breaking of random waves." *Coastal Engineering 1978*, 569-587.

Booij, N., Ris, R. C., and Holthuijsen, L. H. (1999). "A third-generation wave model for coastal regions - 1. Model description and validation." *J Geophys Res-Oceans*, 104(C4), 7649-7666.

Brocchini, M., and Dodd, N. (2008). "Nonlinear shallow water equation modeling for coastal engineering." *J Waterw Port C-Asce*, 134(2), 104-120.

Buckley, M., Lowe, R., and Hansen, J. (2014). "Evaluation of nearshore wave models in steep reef environments." *Ocean Dynamics*, 64(6), 847-862.

Cavaleri, L., Alves, J. H. G. M., Ardhuin, F., Babanin, A., Banner, M., Belibassakis, K., Benoit, M., Donelan, M., Groeneweg, J., Herbers, T. H. C., Hwang, P., Janssen, P. A. E. M., Janssen, T., Lavrenov, I. V., Magne, R., Monbaliu, J., Onorato, M., Polnikov, V., Resio, D., Rogers, W. E., Sheremet, A., McKee Smith, J., Tolman, H. L., van Vledder, G., Wolf, J., and Young, I. (2007). "Wave modelling – The state of the art." *Progress in Oceanography*, 75(4), 603-674.

Crespo, A. J. C., Domínguez, J. M., Rogers, B. D., Gómez-Gesteira, M., Longshaw, S., Canelas, R., Vacondio, R., Barreiro, A., and García-Feal, O. (2015). "DualSPHysics: Open-source parallel CFD solver based on Smoothed Particle Hydrodynamics (SPH)." *Computer Physics Communications*, 187, 204-216.

De Ridder, M. J. D. U. o. T. M. T. R. f. h. r. t. n. u. f. b. a. d. a. e. f. (2018). "Non‐hydrostatic wave modelling of coral reefs with the addition of a porous in‐canopy model."

Gourlay, M. R. (1996). "Wave set-up on coral reefs. 1. Set-up and wave-generated flow on an idealised two dimensional horizontal reef." *Coastal Engineering*, 27(3), 161-193.

Higuera, P., Lara, J. L., and Losada, I. J. (2013). "Simulating coastal engineering processes with OpenFOAM®." *Coastal Engineering*, 71, 119-134.

Hirt, C. W., and Nichols, B. D. (1981). "Volume of fluid (VOF) method for the dynamics of free boundaries." *Journal of Computational Physics*, 39(1), 201-225.

Hofland, B., Chen, X., Altomare, C., and Oosterlo, P. (2017). "Prediction formula for the spectral wave period $T_{m-1,0}$ on mildly sloping shallow foreshores." *Coastal Engineering*, 123(Supplement C), 21-28.

Holthuijsen, L. H., Booij, N., and Herbers, T. H. C. (1989). "A prediction model for stationary, short-crested waves in shallow water with ambient currents." *Coastal Engineering*, 13(1), 23-54.

Jacobsen, N. G., Fuhrman, D. R., and Fredsøe, J. (2012). "A wave generation toolbox for the open-source CFD library: OpenFoam®." 70(9), 1073-1088.

Jasak, H., Jemcov, A., and Tukovic, Z. "OpenFOAM: A C++ library for complex physics simulations." *Proc., International workshop on coupled methods in numerical dynamics*, IUC Dubrovnik, Croatia, 1-20.

Kazolea, M., and Ricchiuto, M. (2018). "On wave breaking for Boussinesq-type models." *Ocean Modelling*, 123, 16-39.

Kirby, J. T., Wei, G., Chen, Q., Kennedy, A. B., and Dalrymple, R. A. (1998). "FUNWAVE 1.0: fully nonlinear Boussinesq wave model-Documentation and user's manual." *research report NO. CACR-98-06*.

Larsen, B. E., and Fuhrman, D. R. (2018). "On the over-production of turbulence beneath surface waves in Reynolds-averaged Navier–Stokes models." *Journal of Fluid Mechanics*, 853, 419-460.

Lashley, C. H., Bricker, J. D., Van der Meer, J., Altomare, C., and Suzuki, T. (*Forthcoming*). "Relative Magnitude of Infragravity Waves at Coastal Dikes with Shallow Foreshores: A Prediction Tool." *Journal of Waterway, Port, Coastal, and Ocean Engineering*.

Lashley, C. H., Roelvink, D., van Dongeren, A., Buckley, M. L., and Lowe, R. J. (2018). "Nonhydrostatic and surfbeat model predictions of extreme wave run-up in fringing reef environments." *Coastal Engineering*, 137, 11-27.

Lowe, R. J., Buckley, M. L., Altomare, C., Rijnsdorp, D. P., Yao, Y., Suzuki, T., and Bricker, J. D. (2019). "Numerical simulations of surf zone wave dynamics using Smoothed Particle Hydrodynamics." *Ocean Modelling*, 144, 101481.

Lynett, P. J., Gately, K., Wilson, R., Montoya, L., Arcas, D., Aytore, B., Bai, Y., Bricker, J. D., Castro, M. J., Cheung, K. F., David, C. G., Dogan, G. G., Escalante, C., González-Vida, J. M., Grilli, S. T., Heitmann, T. W., Horrillo, J., Kânoğlu, U., Kian, R., Kirby, J. T., Li, W., Macías, J., Nicolsky, D. J., Ortega, S., Pampell-Manis, A., Park, Y. S., Roeber, V., Sharghivand, N., Shelby, M., Shi, F., Tehranirad, B., Tolkova, E., Thio, H. K., Velioğlu, D., Yalçıner, A. C., Yamazaki, Y., Zaytsev, A., and Zhang, Y. J. (2017). "Inter-model analysis of tsunami-induced coastal currents." *Ocean Modelling*, 114, 14-32.

Ma, G., Shi, F., and Kirby, J. T. (2012). "Shock-capturing non-hydrostatic model for fully dispersive surface wave processes." *Ocean Modelling*, 43-44, 22-35.

Mase, H., Tamada, T., Yasuda, T., Hedges, T. S., and Reis, M. T. (2013). "Wave Runup and Overtopping at Seawalls Built on Land and in Very Shallow Water." *Journal of Waterway, Port, Coastal, and Ocean Engineering*, 139(5), 346-357.

Nairn, R. B., Roelvink, J., and Southgate, H. N. (1991). "Transition zone width and implications for modelling surfzone hydrodynamics." *Coastal Engineering 1990*, 68-81.

Nwogu, O. (1993). "Alternative Form of Boussinesq Equations for Nearshore Wave Propagation." 119(6), 618-638.

Oosterlo, P., McCall, R., Vuik, V., Hofland, B., van der Meer, J., and Jonkman, S. (2018). "Probabilistic Assessment of Overtopping of Sea Dikes with Foreshores including Infragravity Waves and Morphological Changes: Westkapelle Case Study." *Journal of Marine Science and Engineering*, 6(2), 48.

Rijnsdorp, D. P., Smit, P. B., Zijlema, M., and Reniers, A. J. H. M. (2017). "Efficient non-hydrostatic modelling of 3D wave-induced currents using a subgrid approach." *Ocean Modelling*, 116, 118-133.

Roeber, V., and Cheung, K. "BOSZ (Boussinesq Ocean and Surf Zone model)." *Proc., Proceedings and Results of the 2011 NTHMP Model Benchmarking Workshop, NOAA, Galveston, Texas*.

Roeber, V., Cheung, K. F., and Kobayashi, M. H. (2010). "Shock-capturing Boussinesq-type model for nearshore wave processes." *Coastal Engineering*, 57(4), 407-423.

Roelvink, D., and Costas, S. (2019). "Coupling nearshore and aeolian processes: XBeach and duna process-based models." *Environmental Modelling & Software*, 115, 98-112.

Roelvink, D., McCall, R., Mehvar, S., Nederhoff, K., and Dastgheib, A. (2018). "Improving predictions of swash dynamics in XBeach: The role of groupiness and incident-band runup." *Coastal Engineering*, 134, 103-123.

Roelvink, D., Reniers, A., van Dongeren, A., van Thiel de Vries, J., McCall, R., and Lescinski, J. (2009). "Modelling storm impacts on beaches, dunes and barrier islands." *Coastal Engineering*, 56(11-12), 1133-1152.

Roelvink, J. A. (1993). "Dissipation in random wave groups incident on a beach." *Coastal Engineering*, 19(1), 127-150.

Roelvink, J. A., and Stive, M. J. F. (1989). "Bar‐generating cross‐shore flow mechanisms on a beach." *Journal of Geophysical Research: Oceans (1978–2012)*, 94(C4), 4785-4800.

Romano, A., Bellotti, G., Briganti, R., and Franco, L. (2015). "Uncertainties in the physical modelling of the wave overtopping over a rubble mound breakwater: The role of the seeding number and of the test duration." *Coastal Engineering*, 103, 15-21.

Sierra, J. P., González-Marco, D., Mestres, M., Gironella, X., Oliveira, T. C. A., Cáceres, I., and Mösso, C. (2010). "Numerical model for wave overtopping and transmission through permeable coastal structures." *Environmental Modelling & Software*, 25(12), 1897-1904.

Simarro, G., Orfila, A., and Galan, A. J. C. e. (2013). "Linear shoaling in Boussinesq-type wave propagation models." 80, 100-106.

Smit, P., Stelling, G., Roelvink, J., Van Thiel de Vries, J., McCall, R., Van Dongeren, A., Zwinkels, C., and Jacobs, R. (2010). "XBeach: Non-hydrostatic model: Validation, verification and model description." *Delft Univ. Technol*.

Smit, P., Zijlema, M., and Stelling, G. (2013). "Depth-induced wave breaking in a non-hydrostatic, near-shore wave model." *Coastal Engineering*, 76, 1-16.

Smith, J. M., Sherlock, A. R., and Resio, D. T. (2001). "STWAVE: Steady-state spectral wave model user's manual for STWAVE, Version 3.0." ENGINEER RESEARCH AND DEVELOPMENT CENTER VICKSBURG MS COASTAL AND HYDRAULICSLAB.

Smith, R. A. E., Bates, P. D., and Hayes, C. (2012). "Evaluation of a coastal flood inundation model using hard and soft data." *Environmental Modelling & Software*, 30, 35-46.

St-Germain, P., Nistor, I., Readshaw, J., and Lamont, G. (2014). "NUMERICAL MODELING OF COASTAL DIKE OVERTOPPING USING SPH AND NON-HYDROSTATIC NLSW EQUATIONS." *2014*(34).

Stelling, G., and Zijlema, M. (2003). "An accurate and efficient finite-difference algorithm for non-hydrostatic free-surface flow with application to wave propagation." *International Journal for Numerical Methods in Fluids*, 43(1), 1-23.

Stockdon, H. F., Holman, R. A., Howd, P. A., and Sallenger, A. H. (2006). "Empirical parameterization of setup, swash, and runup." *Coastal Engineering*, 53(7), 573-588.

Suzuki, T., Altomare, C., Veale, W., Verwaest, T., Trouw, K., Troch, P., and Zijlema, M. (2017). "Efficient and robust wave overtopping estimation for impermeable coastal structures in shallow foreshores using SWASH." *Coastal Engineering*, 122(Supplement C), 108-123.

Svendsen, I. A. (1984). "Wave heights and set-up in a surf zone." *Coastal Engineering*, 8(4), 303-329.

Van der Meer, J., Allsop, N., Bruce, T., De Rouck, J., Kortenhaus, A., Pullen, T., Schuttrumpf, H., Troch, P., and Zanuttigh, B. (2018). "EurOtop 2018: Manual on wave overtopping of sea defences and related structures. An overtopping manual largely based on European research, but for worldwide application." Retrieved from www. overtopping-manual. com.

Van Gent, M. (1999). "Wave run-up and wave overtopping for double peaked wave energy spectra." *H3351*.

Vanneste, D. F. A., Altomare, C., Suzuki, T., Troch, P., and Verwaest, T. (2014). "COMPARISON OF NUMERICAL MODELS FOR WAVE OVERTOPPING AND IMPACT ON A SEA WALL." *2014*(34).

Verbrugghe, T., Domínguez, J. M., Crespo, A. J. C., Altomare, C., Stratigaki, V., Troch, P., and Kortenhaus, A. (2018). "Coupling methodology for smoothed particle hydrodynamics modelling of non-linear wave-structure interactions." *Coastal Engineering*, 138, 184-198.

Vuik, V., Jonkman, S. N., Borsje, B. W., and Suzuki, T. (2016). "Nature-based flood protection: The efficiency of vegetated foreshores for reducing wave loads on coastal dikes." *Coastal Engineering*, 116(Supplement C), 42-56.

Vyzikas, T., and Greaves, D. (2018). "Numerical Modelling." *Wave and Tidal Energy*, 289-363.

Warren, I. R., and Bach, H. K. (1992). "MIKE 21: a modelling system for estuaries, coastal waters and seas." *Environmental Software*, 7(4), 229-240.

Yang, S. L., Shi, B. W., Bouma, T. J., Ysebaert, T., and Luo, X. X. (2012). "Wave Attenuation at a Salt Marsh Margin: A Case Study of an Exposed Coast on the Yangtze Estuary." *Estuaries and Coasts*, 35(1), 169-182.

Zhang, S.-j., Zhu, L.-s., and Zou, K. J. C. O. E. (2019). "A Comparative Study of Numerical Models for Wave Propagation and Setup on Steep Coral Reefs." 33(4), 424-435.

Zijlema, M., Stelling, G., and Smit, P. (2011). "SWASH: An operational public domain code for simulating wave fields and rapidly varied flows in coastal waters." *Coastal Engineering*, 58(10), 992-1012.

Zijlema, M., and Stelling, G. S. (2008). "Efficient computation of surf zone waves using the nonlinear shallow water equations with non-hydrostatic pressure." *Coastal Engineering*, 55(10), 780-790.